\documentclass[aip,graphicx,reprint]{revtex4-2}
\usepackage{mhchem, braket, graphicx, subcaption}
\usepackage{goldstone}
\usepackage[offset=1.2em, sep=0pt]{simpler-wick} 
\usepackage{enumitem} 
\newcommand{\cm}{\ensuremath{\text{cm}^{-1}}}

\begin{document}
	\title{Assessing the Orbital-Optimized Unitary Ansatz for Density Cumulant Theory}

\author{Jonathon P. Misiewicz}
\author{Justin M. Turney}
\author{Henry F. Schaefer III}
\email{ccq@uga.edu}
\affiliation
{Center for Computational Quantum Chemistry, University of Georgia, Athens, Georgia 30602, United States}
\author{Alexander Yu. Sokolov}
\affiliation
{Department of Chemistry and Biochemistry, The Ohio State University, Columbus, Ohio 43210, United States}

	\date{\today}

	\begin{abstract}
The previously proposed ansatz for density cumulant theory that combines orbital-optimization and a parameterization of the 2-electron reduced density matrix cumulant in terms of unitary coupled cluster amplitudes (OUDCT) is carefully examined. Formally, we elucidate the relationship between OUDCT and orbital-optimized unitary coupled cluster theory and show the existence of near-zero denominators in the stationarity conditions for both the exact and some approximate OUDCT methods. We implement methods of the OUDCT ansatz restricted to double excitations for numerical study, up to the fifth commutator in the Baker-Campbell-Hausdorff expansion. We find that methods derived from the ansatz beyond the previously known ODC-12 method tend to be less accurate for equilibrium properties and less reliable when attempting to describe \ce{H2} dissociation. New developments are needed to formulate more accurate DCT variants.
	\end{abstract}

	\maketitle

\section{Introduction}
ODC-12 is the most successful method to date of the density cumulant theory\cite{Kutzelnigg:2006p171101, Sokolov:2013p024107, Sokolov:2013p204110} (DCT) family of electronic structure methods. For a system of $o$ occupied orbitals and $v$ virtual orbitals, ODC-12 has the $\mathcal{O}(o^2v^4)$ scaling of coupled cluster with singles and double excitations\cite{Shavitt:2009, Bartlett:2007p291} (CCSD) but is consistently more accurate.\cite{Copan:2014p2389, Sokolov:2013p204110, Wang:2016p4833} It has a simple, inexpensive analytic gradient theory.\cite{Sokolov:2013p204110} It tolerates multireference effects that leave CCSD qualitatively incorrect.\cite{Mullinax_2015} For these reasons, there has been interest in extending the success of ODC-12 both to achieve greater accuracy for weakly correlated molecules, and to develop a method able to treat multiconfigurational molecules.\cite{Sokolov:2014p074111, Mullinax_2015, Cioslowski:2019p4862}
	
To date, there has been only one published proposal of density cumulant theory methods going beyond ODC-12 in accuracy.\cite{Sokolov:2014p074111} Reference \citenum{Sokolov:2014p074111} introduced a formally exact ansatz for density cumulant theory and proposed that approximating it may yield the desired improvements to ODC-12. As a proof-of-concept, the authors implemented and benchmarked the ODC-13 method, which adds terms to ODC-12 and is derived from the aforementioned ansatz.

Unfortunately, the ODC-13 method did not improve on the success of the simpler ODC-12. The authors of Reference \citenum{Sokolov:2014p074111} reported that ODC-13 was less accurate in the weakly correlated regime, as determined by comparison against experimental bond lengths and vibrational frequencies for diatomic molecules. The accuracy of the method across various correlation strengths was assessed by \ce{H2} dissociation. For this system, ODC-13 was less accurate than ODC-12 past 0.9 \AA\ and could not be converged beyond 1.3 \AA. Reference \citenum{Sokolov:2014p074111} observed that a particular exact relationship between two key intermediates of the theory, the 1-electron reduced density matrix (1RDM) and the 2-electron reduced density matrix (2RDM), was not satisfied in ODC-13. The authors suggested that violating that relationship might have caused the ``unsatisfactory'' performance of ODC-13.

Additionally, the authors of Reference \citenum{Sokolov:2014p074111} proposed but did not implement an alternative scheme to approximate their ansatz where that relationship is obeyed. This approximation consists of truncating a Baker-Campbell-Hausdorff (BCH) expansion to a finite number of commutators to obtain one part of the 2RDM while maintaining the aforementioned exact relation between the 1RDM and the 2RDM to determine the rest. As there have been no further studies of any post-ODC-12 methods of this ansatz, it remains untested whether the failure of ODC-13 can be attributed to violations of this relationship or whether the performance of ODC-13 indicates a more general complication in working with the DCT ansatz advanced in Reference \citenum{Sokolov:2014p074111}.

In this article, we study truncations of the orbital-optimized unitary coupled cluster ansatz for DCT (OUDCT) proposed in Reference \citenum{Sokolov:2014p074111}. We begin in Section \ref{sec:oudct} with a thorough review of the equations of the OUDCT ansatz, which are scattered across multiple papers.\cite{Kutzelnigg:2006p171101, Sokolov:2013p024107, Sokolov:2013p204110, Sokolov:2014p074111} During this review, two new \textit{formal} questions about the ansatz arise, namely:
\begin{enumerate}
	\item The residual equations for the OUDCT stationarity conditions contain terms with near-zero denominators, which all vanish in ODC-12. Do these vanish in the exact OUDCT theory?
	\item The similar orbital-optimized variational unitary coupled cluster method can also be approximated by truncations of the BCH expansion to a finite number of commutators. What is the relationship between approximations of that method and OUDCT, truncated at the same degree?
\end{enumerate}

\noindent We then turn our attention to numerical studies of the performance of low-degree OUDCT truncations, with only double excitations. The truncation at degree two is the aforementioned ODC-12 method and is our baseline for both accuracy and degree of truncation. After discussing our implementation of the methods in Section \ref{sec:implement}, we perform numerical studies in Section \ref{sec:numerical}. We investigate:
\begin{enumerate}[resume]
	\item Do higher-degree truncations of the OUDCT ansatz improve the accuracy for \ce{H2}? \ce{H2} is important both as a case where effects of triple and higher-rank cluster operators do not exist, and as a model of variable correlation strength.
	\item Do higher-degree truncations of the OUDCT ansatz, restricted to doubles, improve the accuracy for systems with more than two electrons, where triples effects may be important?
\end{enumerate}

The results of our investigation lead us to conclude that OUDCT ansatz truncations that include more commutators than ODC-12, up to five, improve upon ODC-13 for \ce{H2} dissociation, but are still inferior to ODC-12 for moderate bond stretching. For equilibrium properties of weakly correlated molecules with more than two electrons, OUDCT approximations will not improve upon ODC-12 unless unitary cluster operators beyond doubles are accounted for. Further, treating triple unitary cluster operators to four or more commutators in the BCH expansion will lead to singularities in the theory.

\section{The Orbital-Optimized Unitary Density Cumulant Theory Ansatz}
\label{sec:oudct}
This section provides a self-contained exposition of DCT and the OUDCT ansatz in particular, starting from an understanding of electron correlation at the level of Shavitt and Bartlett's text\cite{Shavitt:2009} and a loose acquaintance with reduced density matrix (RDM) theory.\cite{Ciosloswki2000, ACP134} Section \ref{sec:abstract-dct} derives the theoretical essentials of DCT, bypassing the intermediates $\kappa$ and $\tau$ of Reference \citenum{Kutzelnigg:2006p171101}. The degeneracies in the cumulant partial trace discussed in that section have not been discussed previously. Section \ref{sec:ucc} and Section \ref{sec:udct} derive the Variational Unitary Coupled Cluster (VUCC) and Unitary DCT (UDCT) ans{\"a}tze. The latter should be compared with Reference \citenum{Sokolov:2014p074111}. Section \ref{sec:oo} discusses the addition of orbital optimization to the VUCC and UDCT ans{\"a}tze to produce OVUCC and OUDCT, focusing on the implications and advantages of doing so. Orbital optimization was added to DCT in Reference \citenum{Sokolov:2013p204110}. In this section, we also discuss the possibility of singularities in the cumulant update equations, as these depend not only on the cumulant parameterization but also on the orbitals. Finally, Section \ref{sec:udct-ucc-compare} formally analyzes the difference between UDCT and UCC truncated at the same degree.

Throughout this section, we use primed indices to denote a quantity that must be computed in the basis of natural orbitals.

\subsection{Abstract Density Cumulant Theory}
\label{sec:abstract-dct}

We begin by writing the Hamiltonian in second-quantized form

\begin{equation}
	\label{eq:H}
	\hat{H} = h_p^q a^p_q + \frac{1}{4} \bar{g}_{pq}^{rs} a^{pq}_{rs}
\end{equation}

\noindent where $h_p^q$ is the standard one-electron integral, $\bra{\phi_p} \hat{h} \ket{\phi_q}$, and $\bar{g}_{pq}^{rs}$ is the antisymmetrized electron repulsion integral, $\bra{pq} \ket{rs}$. We use the notation introduced by Kutzelnigg in Reference \citenum{Kutzelnigg:1982p3081} for writing the vacuum-normal, particle-conserving second quantized fermionic operators ($a^p_q = a_p^\dagger a_q$ and $a^{pq}_{rs} = a_p^\dagger a_q^\dagger a_s a_r$), and we also use the Einstein summation convention throughout this article.

It follows from \eqref{eq:H} that the energy expectation value of any normalized wavefunction $\Psi$ may be written as

\begin{equation}
	\label{eq:rdm-energy}
	E = h_p^q \gamma^p_q + \frac{1}{4} \bar{g}_{pq}^{rs} \gamma^{pq}_{rs}
\end{equation}

\noindent respectively defining the 1-electron RDM (1RDM) and 2-electron RDM (2RDM) with

\begin{equation}
	\label{eq:1rdm}
	\gamma^p_q = \bra{\Psi} a^p_q \ket{\Psi}
\end{equation}

\noindent and

\begin{equation}
	\label{eq:2rdm}
	\gamma^{pq}_{rs} = \bra{\Psi} a^{pq}_{rs} \ket{\Psi} \quad.
\end{equation}

\noindent For exact wavefunctions, $\gamma^{pq}_{rs}$ is multiplicatively separable, not additively separable, i.e., not size-consistent.\cite{Misiewicz:2020pX} We may decompose it into size-consistent tensors with

\begin{equation}
	\label{eq:2rdm-decomp}
	\gamma^{pq}_{rs} = \lambda^{pq}_{rs} + \gamma^p_r\gamma^q_s - \gamma^p_s \gamma^q_r \quad .
\end{equation}

\noindent It may be checked manually that if $\gamma^{pq}_{rs}$ is multiplicatively separable, $\lambda^{pq}_{rs}$ must be zero.\cite{Misiewicz:2020pX} These size-consistent tensors are called RDM cumulants and denoted with $\lambda$.\cite{Misiewicz:2020pX, Hanauer:2012p50, Mazziotti:2011p244, Mazziotti:1998p419, Kutzelnigg:1999p2800, Ziesche:2000p33, Piris:2014p1169, Torre:2003p127} Because the 1RDM equals its cumulant, we shall usually refer to it with $\gamma$, as done in \eqref{eq:2rdm-decomp}. When no superscripts or subscripts specify the rank of the tensor, $\lambda$ shall refer to the 2RDM cumulant, and $\gamma$ shall refer to the 1RDM.

Substituting \eqref{eq:2rdm-decomp} into \eqref{eq:rdm-energy} and using antisymmetry of $\bar{g}$ gives

\begin{equation}
	\label{eq:cumulant-energy}
	E = (h_p^q + \frac{1}{2} \bar{g}_{pr}^{qs} \gamma_s^r)  \gamma^p_q + \frac{1}{4} \bar{g}_{pq}^{rs} \lambda^{pq}_{rs} \quad .
\end{equation}

\noindent This is an exact functional of $\gamma$ and $\lambda$. To find the exact ground-state energy, we want to minimize this functional over the set of $\gamma$ and $\lambda$ possible, given the definitions of \eqref{eq:1rdm}, \eqref{eq:2rdm}, and \eqref{eq:2rdm-decomp}. A pair of $\gamma$ and $\lambda$ consistent with those equations is said to be pure $n$-representable.\cite{Coleman:1963p668, Schilling:2018p231102, Klyachko:2006p72, Altunbulak:2008p287, Coleman:1978p67, Mazziotti:2016p032516} However, the set of pure $n$-representable $\gamma$ and $\lambda$ has a complicated structure, and there is no known parameterization that is necessary, sufficient, and computationally efficient. Accordingly, our strategy will be to take a parameterization that is necessary and sufficient, approximate it for computational efficiency, and vary the amplitudes until the derivative of the energy functional is zero.

At first sight, we need to parameterize both $\gamma$ and $\lambda$. However, for a given $\lambda$, the set of $\gamma$ consistent with it is strongly constrained. While $\gamma$ is not a function of $\lambda$ (we will construct a counterexample later in this subsection), the set of possible $\gamma$ is discrete for all but exceptional $\lambda$. This enables us to use implicit differentiation to treat \eqref{eq:cumulant-energy} as a function of the $\lambda$ parameters alone for differentiation purposes, so we need only parameterize $\lambda$.

We will begin by constraining the set of $\gamma$ that are consistent with a given $\lambda$.\cite{Kutzelnigg:2006p171101} For an $n$-electron system,

\begin{equation}
	\label{eq:rdm-partial-trace}
	\gamma^{pr}_{qr} = (n-1) \gamma^p_q
\end{equation}

\noindent and

\begin{equation}
	\label{eq:rdm-partial-trace2}
	\gamma^{p}_{p} = n \quad .
\end{equation}

\noindent (Equations \eqref{eq:rdm-partial-trace} and \eqref{eq:rdm-partial-trace2} are easily proven by expanding $\Psi$ from \eqref{eq:1rdm} and \eqref{eq:2rdm} in terms of Slater determinants.) Inserting \eqref{eq:2rdm-decomp} in \eqref{eq:rdm-partial-trace} yields, through straightforward algebra and an invocation of \eqref{eq:rdm-partial-trace2}:

\begin{equation}
	\label{eq:rdmc-partial-trace}
	d^p_q = (\gamma^2 - \gamma)^p_q
\end{equation}

\noindent where

\begin{equation}
	\label{eq:def-d}
	d^p_q = \lambda^{pr}_{qr} \quad .
\end{equation}

\noindent The matrix $d$ is quadratic in the matrix $\gamma$. The set of $\gamma$ consistent with \eqref{eq:rdmc-partial-trace} for a given $d$ may be characterized as follows:
If $\gamma$ is consistent with Equation \eqref{eq:rdmc-partial-trace}, then express Equation \eqref{eq:rdmc-partial-trace} in an eigenbasis of $\gamma$ where eigenvector $v_p$ has eigenvalue $\gamma_{p^\prime}$. The eigenvectors are called the natural spin-orbitals, and the eigenvalues are the natural spin-orbital occupation numbers. Then the right-hand side of Equation \eqref{eq:rdmc-partial-trace} is a diagonal matrix with entries $\gamma_{p^\prime}^2 - \gamma_{p^\prime}$. It follows that each eigenvector $v_p$ is also an eigenvector of $d$ with eigenvalues $\Delta_{p^\prime} = \gamma_{p^\prime}^2 - \gamma_{p^\prime}$. This may be solved to yield
\begin{equation}
	\label{eq:d-to-gamma-eval}
	\gamma_{p^\prime} = \frac{1 \pm \sqrt{1+4\Delta_{p^\prime}}}{2} \quad .
\end{equation}

\begin{figure}
	\caption{The natural spin-orbital (NSO) occupation number $\gamma_{p^\prime}$ as a multi-valued function of the corresponding eigenvalue in the partial trace of the 2RDM cumulant, $d_{p^\prime}$. In general, a $d_{p^\prime}$ eigenvalue is consistent with two possible occupation numbers, one suggesting an occupied orbital and the other suggesting a virtual orbital.}
	\includegraphics[width=0.8\linewidth]{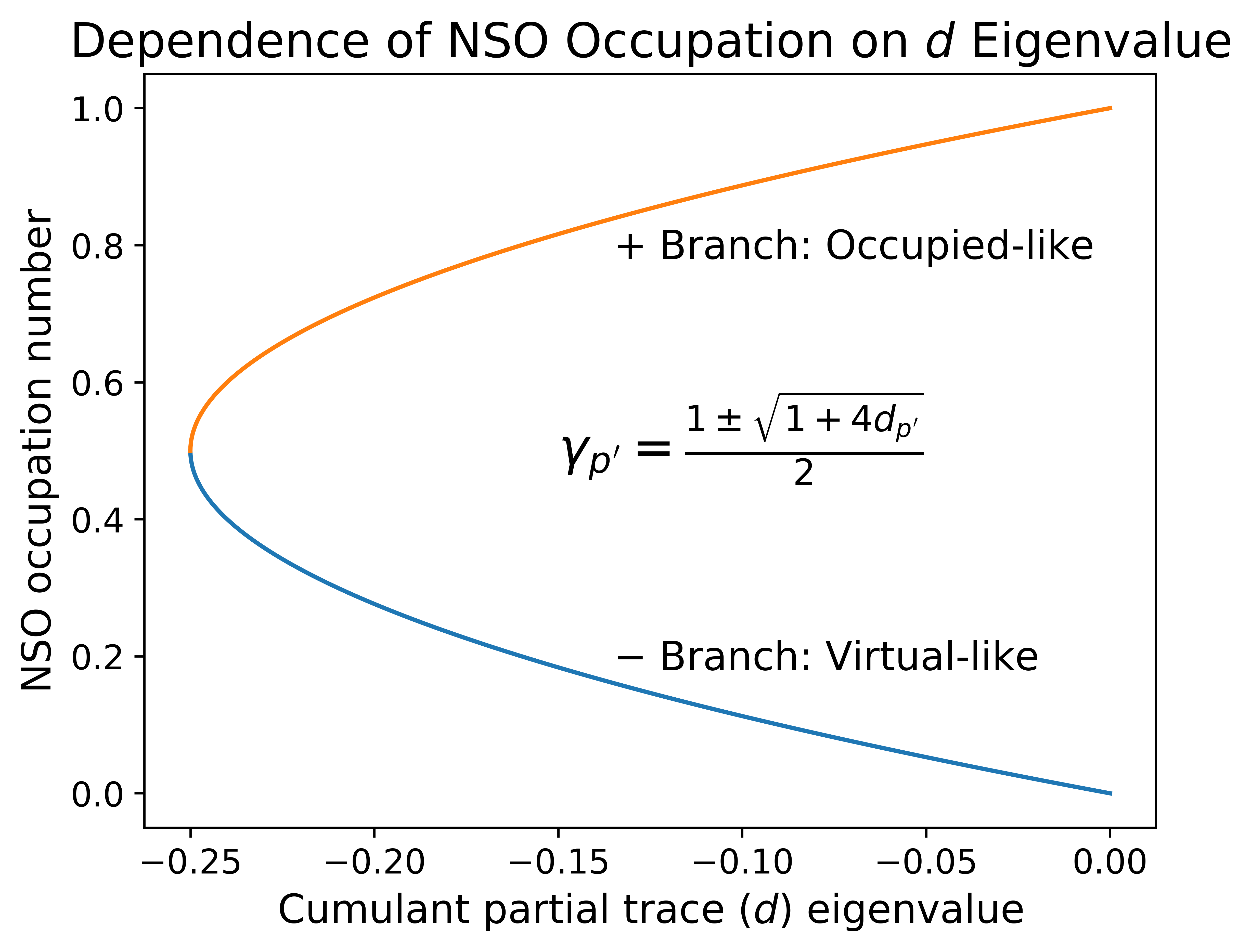}
	\label{fig:d-gamma}
\end{figure}

\noindent Choosing the $+$ sign is consistent with $\gamma_{p^\prime} \geq \frac{1}{2}$, and choosing the $-$ sign is consistent with $\gamma_{p^\prime} \leq \frac{1}{2}$. These choices are illustrated in Figure \ref{fig:d-gamma}. Therefore, for all $\gamma$ consistent with Equation \eqref{eq:rdmc-partial-trace}, it is necessary that there exist some eigenbasis of $d$ that is also an eigenbasis of $\gamma$ with eigenvalues from \eqref{eq:d-to-gamma-eval}. It is clear that the existence of such a $d$ eigenbasis is sufficient to satisfy \eqref{eq:rdmc-partial-trace} in the chosen eigenbasis, and thus in any basis. Therefore, the set of $\gamma$ so constructed from $d$ is precisely the set of solutions to \eqref{eq:rdmc-partial-trace}. Equations \eqref{eq:rdmc-partial-trace} and \eqref{eq:def-d} are merely necessary for $\gamma$ and $\lambda$ to be pure $n$-representable, not sufficient, but we shall not need sufficiency.

Equivalently, solutions to \eqref{eq:rdmc-partial-trace} take the form
\begin{equation}
\label{eq:soln-space}
\gamma = U \begin{bmatrix}
	\frac{1 + \sqrt{1+4\Delta_o}}{2} & 0\\
	0 & \frac{1 - \sqrt{1+4\Delta_v}}{2}
\end{bmatrix} U^{-1}
\end{equation}

\noindent where the matrix $U$ is some matrix of eigenvectors of $d$,

\begin{equation}
	\label{eq:def-u}
	\Delta = U^{-1} \: d \; U
\end{equation}

\noindent and $\Delta_o$ and $\Delta_v$ are the occupied and virtual blocks of $\Delta$, the matrix of eigenvalues of $d$. A natural orbital taking the $+$ solution of \eqref{eq:d-to-gamma-eval} is equivalent to it being a column of $U$ that is multiplied against the (occupied) $+$ block of \eqref{eq:soln-space}, and analogously for the (virtual) $-$ block.

Depending on the eigenvalue structure of $d$, the set of possible $\gamma$ from \eqref{eq:soln-space} may be either discrete or continuous. If there are no degeneracies in the $d$ matrix, then the eigenvectors of $d$ are unambiguous, and so are the eigenvectors of $\gamma$. It remains only to choose whether a given natural spin-orbital occupation number should take the $+$ sign (occupied-like) or the $-$ sign (virtual-like) in \eqref{eq:d-to-gamma-eval}. If the goal is to approximate some electronic state, these can be chosen by comparing the natural spin-orbitals to those of another approximation to the electronic state and choosing the signs to mimic the occupation numbers of the other approximation.\cite{Nooijen:1999p8356, Herbert_2007} This other approximation may be a Hartree--Fock computation or a previous solution to \eqref{eq:rdmc-partial-trace}.

However, let us suppose that there is a degeneracy in the $d$ matrix. There remains the discrete freedom in how many of an eigenspace's eigenvectors take the $+$ sign and how many $-$, but there is also a new continuous freedom in partitioning the eigenspaces into $+$ and $-$ eigenvectors. A concrete example of this is a single determinant wavefunction, which has $\lambda = 0$. Then \eqref{eq:d-to-gamma-eval} implies that the natural orbitals all have occupation number $0$ or $1$, but the requirement that the wavefunction is a Slater determinant does not determine which orbitals are occupied and which are virtual. Instead, there is a continuous set of possible Slater determinants, as is well-known from the continuous Hartree--Fock problem. This also shows that a pure $n$-representable $\lambda$ need not uniquely determine the corresponding $\gamma$. This continuous freedom from degeneracies was not considered by previous research that derived \eqref{eq:d-to-gamma-eval}.\cite{Kutzelnigg:2006p171101, Nooijen:1999p8356, Herbert_2007}

Let us use these results to write $\gamma$ as a (continuously) differentiable implicit function of $\lambda$ about some neighborhood of a starting solution to \eqref{eq:rdmc-partial-trace}. If we cast $\gamma$ in the basis of natural orbitals, we find that

\begin{equation}
	\label{eq:gamma-lambda-deriv}
	\frac{\partial}{\partial d^{p^\prime}_{q^\prime}} \gamma^{p^\prime}_{q^\prime} = \frac{1}{\gamma_{p^\prime} + \gamma_{q^\prime} - 1} \quad .
\end{equation}

\noindent This equation has singularities if $\gamma_{p^\prime} + \gamma_{q^\prime} = 1$, which are precisely the cases where $\frac{d}{d \gamma} (\gamma^2 - \gamma)$ fails to be invertible. $\frac{d}{d \gamma} (\gamma^2 - \gamma)$ is the matrix $A_x$ in Theorem 9.28 of Reference \citenum{walterrudin1976}, so the hypotheses of the implicit function theorem are not satisfied in this case, and implicit differentiation fails.

We may interpret these singularities via the discussion regarding solutions to \eqref{eq:rdmc-partial-trace}. If $p^\prime = q^\prime$ and $\gamma_{p^\prime} + \gamma_{q^\prime} = 1$, then orbital $p^\prime$ is half-occupied, and either choice of sign in \eqref{eq:d-to-gamma-eval} gives the same result. It is undetermined from \eqref{eq:rdmc-partial-trace} whether a change in $d$ will cause the occupation number to take the $+$ sign and be slightly more occupied, or to take the $-$ sign and be slightly more virtual.

If $p^\prime \neq q^\prime$ and $\gamma_{p^\prime} + \gamma_{q}^\prime = 1$, then $p^\prime$ and $q^\prime$ have a common $d$ eigenvalue by \eqref{eq:rdmc-partial-trace}, and we must decide how to split their degeneracy in the $d$ matrix into an occupied and a virtual orbital. Any slight perturbation of the $d$ matrix may break the degeneracy, causing unpredictable changes in how the orbital spaces break into an occupied and a virtual orbital. This costs differentiability, because the eigenvectors are not even continuous with respect to changes in $d$.

We may now use \eqref{eq:gamma-lambda-deriv} in conjunction with \eqref{eq:cumulant-energy} and \eqref{eq:def-d} to minimize the energy with respect to yet unspecified cumulant parameters. We can explicitly write the derivative of the energy with respect to cumulant parameter $t$ as 

\begin{equation}
	\label{eq:amplitude-gradient}
	\frac{\partial E}{\partial t} = \tilde{F}_p^q \frac{\partial d^p_q}{\partial t} + \bar{g}_{pq}^{rs} \frac{\partial \lambda^{pq}_{rs}}{\partial t}
\end{equation}

\noindent defining

\begin{equation}
	\label{eq:gen-fock}
	\tilde{F}_{p^\prime}^{q^\prime} = \frac{h_{p^\prime}^{q^\prime} + \bar{g}_{p^\prime r^\prime}^{q^\prime s^\prime} \gamma^{r^\prime}_{s^\prime}}{n_{p^\prime} + n_{q^\prime} - 1}
\end{equation}

\noindent which must first be computed in the basis of natural spin-orbitals before being transformed back to the original orbital basis in which the cumulant was constructed.\cite{Sokolov:2013p024107} It remains only to choose a cumulant parameterization.

We note that the formula $\gamma = \kappa + \tau$ used in the original DCT paper\cite{Kutzelnigg:2006p171101} has been entirely eliminated in this presentation. Previous work\cite{Sokolov:2018p204113} also presented formulas without this decomposition, but did not discuss them in detail. Contrary to the claims of Reference \citenum{Kutzelnigg:2006p171101}, $\kappa$ is \textit{not} a variable independent of $\lambda$, as the two strongly constrain each other. However, all previous DCT numerical studies\cite{Simmonett_2010, Sokolov:2012p054105, Sokolov:2013p024107, Sokolov:2013p204110, Copan:2014p2389, Sokolov:2014p074111, Mullinax_2015, Wang:2016p4833, Copan:2018p4097, Peng:2019p1840} obeyed this constraint. The new constraint is discussed in Appendix \ref{sec:kappa-tau}.

\subsection{Variational Unitary Coupled Cluster Ansatz}
\label{sec:ucc}
Assume that any wavefunction, $\Psi$, may be written as

\begin{equation}
	\label{eq:unitary}
	\ket{\Psi} = \exp(T - T^\dagger) \ket{\Phi}
\end{equation}

\noindent for a reference determinant $\ket{\Phi}$ where

\begin{equation}
	\label{eq:def-big-t}
	T = T_1 + T_2 + ...
\end{equation}

\noindent and

\begin{equation}
	\label{eq:def-t}
	T_n = \left(\frac{1}{n!} \right)^2 t^{ij...}_{ab...} a_{ij...}^{ab...} \quad .
\end{equation}

\noindent In other words, it is assumed that the unitary coupled cluster (UCC) ansatz\cite{Cooper:2010p234102, Evangelista:2011p224102, Chen:2012p014108, Bartlett:1989p133, Taube:2006p3393, Kutzelnigg:1991p349, Kutzelnigg_2010, Kutzelnigg_1977} is exact. The validity of this assumption has been studied by Evangelista, Chan, and Scuseria.\cite{Evangelista:2019p244112}

The energy expectation value of this wavefunction is given by \eqref{eq:rdm-energy}, where the RDM formulas \eqref{eq:1rdm} and \eqref{eq:2rdm} may be written as functions of the amplitudes $t$:

\begin{equation}
	\label{eq:unitary-1pdm}
	\gamma^{p}_{q}(t) = \bra{\Phi} \exp(T^\dagger - T) a^{p}_{q} \exp(T-T^\dagger) \ket{\Phi}
\end{equation}

\noindent and

\begin{equation}
	\label{eq:unitary-2pdm}
	\gamma^{pq}_{rs}(t) = \bra{\Phi} \exp(T^\dagger - T) a^{pq}_{rs} \exp(T-T^\dagger) \ket{\Phi} \quad .
\end{equation}

\noindent By using the Baker--Campbell--Hausdorff expansion, \eqref{eq:unitary-1pdm} and \eqref{eq:unitary-2pdm} may be written as

\begin{equation}
	\label{eq:bch-1rdm}
	\gamma^{p}_{q}(t) = \sum\limits_{n = 0}^\infty \frac{1}{n!} \bra{\Phi} [\cdot, T - T^\dagger]^n(a^{p}_{q}) \ket{\Phi}
\end{equation}

and 

\begin{equation}
	\label{eq:bch-2rdm}
	\gamma^{pq}_{rs}(t) = \sum\limits_{n = 0}^\infty \frac{1}{n!} \bra{\Phi} [\cdot, T - T^\dagger]^n(a^{pq}_{rs}) \ket{\Phi}
\end{equation}

\noindent where the function $[\cdot, T](H)$ sends $H$ to $[H, T]$.

The variational unitary coupled cluster (VUCC) ansatz consists of approximating the functions \eqref{eq:bch-1rdm} and \eqref{eq:bch-2rdm}, using those approximations in \eqref{eq:rdm-energy} to construct an approximate energy function of the amplitudes $t$, and taking the energy as the variational minimum of that function. This is equivalent to the more usual definition, where the energy function is defined directly as \noindent

\begin{equation}
	\label{eq:bch-energy}
	E(t) = \sum\limits_{n = 0}^\infty \frac{1}{n!} \bra{\Phi} [\cdot, T - T^\dagger]^n(H) \ket{\Phi}
\end{equation}

\noindent but using RDM intermediates will facilitate comparison with DCT.

\subsection{Unitary Density Cumulant Theory}
\label{sec:udct}

From \eqref{eq:2rdm-decomp}, \eqref{eq:bch-1rdm} and \eqref{eq:bch-2rdm}, we immediately have an exact function from the amplitudes $t$ to $\lambda$. Furthermore, this parameterizes only pure $n$-representable cumulants, and if the UCC ansatz is exact, this parameterizes all pure $n$-representable cumulants. We can thus approximate the map from the $t$ amplitudes to $\lambda$ and use density cumulant theory as developed in Section \ref{sec:abstract-dct} to approximate \eqref{eq:1rdm} and \eqref{eq:2rdm} and perform the variational unitary coupled cluster of Section \ref{sec:ucc}. The only source of error is how we approximate the map from $t$ amplitudes to the cumulant.

Constructing the cumulant function by inserting \eqref{eq:bch-1rdm} and \eqref{eq:bch-2rdm} into \eqref{eq:2rdm-decomp} will lead to a large cancellation of terms. We can instead equate the connected terms on both sides of \eqref{eq:bch-2rdm}.\cite{Sokolov:2014p074111} This is valid because that is the only way to divide the terms of \eqref{eq:bch-2rdm} into pieces with the additive separability structure of \eqref{eq:2rdm-decomp}. Every connected term must be assigned to the cumulant because it cannot arise as a product of disconnected pieces. No disconnected term can be assigned to the cumulant because by the linear independence of monomials in any variables (here the $t$ amplitudes), the cumulant would not be zero as a polynomial in the amplitudes if its orbitals correspond to independent subsystems. Doing this yields the exact relation:\cite{Sokolov:2014p074111}

\begin{equation}
	\label{eq:bch-2rdmc}
	\lambda^{pq}_{rs}(t) = \sum\limits_{n = 0}^\infty \frac{1}{n!} \bra{\Phi} [\cdot, T - T^\dagger]^n(a^{pq}_{rs}) \ket{\Phi}_C \quad .
\end{equation}

Let us make a few observations about these equations.
\begin{enumerate}
	\item It is natural to approximate \eqref{eq:bch-2rdmc} by truncating its Taylor series expansion at some degree in the cluster operators $T$. These degrees in $T$ are what Reference \citenum{Sokolov:2014p074111} meant by orders in perturbation theory. Although methods of the ansatz can be analyzed in terms of the terms produced upon M{\o}ller-Plesset partitioning of the molecular Hamiltonian (MPPT),\cite{Sokolov:2013p024107, Sokolov:2013p204110} and MPPT played a prominent role in the derivation of the cumulant approximation of ODC-12,\cite{Kutzelnigg:2004p7350, Kutzelnigg:2006p171101} MPPT is not necessary to formulate the ansatz.
	\item All DCT publications\cite{Kutzelnigg:2006p171101, Simmonett_2010, Sokolov:2012p054105, Sokolov:2013p024107, Sokolov:2013p204110, Copan:2014p2389, Mullinax_2015, Wang:2016p4833, Copan:2018p4097, Peng:2019p1840} excluding Reference \citenum{Sokolov:2014p074111} parameterized the cumulant in terms of parameters $t^{ij}_{ab}$ satisfying $\lambda^{ij}_{ab} = t^{ij}_{ab}$. This can be derived by approximating the UDCT cumulant parameterization \eqref{eq:bch-2rdmc} to two commutators and $T = T_2$. The equation $\lambda^{ij}_{ab} = t^{ij}_{ab}$ can alternatively be interpreted as identifying the parameters as cumulant elements, hence why the parameters were written as $\lambda^{ij}_{ab}$ in many DCT publications.\cite{Kutzelnigg:2006p171101, Simmonett_2010, Sokolov:2012p054105, Sokolov:2013p024107, Sokolov:2013p204110, Wang:2016p4833} If the cluster operator includes $T_2$ and $T_4$ and is truncated after two commutators, or if it includes $T_2$ but is truncated after three commutators, approximations to \eqref{eq:bch-2rdmc} will no longer be consistent with $\lambda^{ij}_{ab} = t^{ij}_{ab}$, and the amplitudes can no longer be identified with cumulant elements. For the exact ansatz of Reference \citenum{Sokolov:2014p074111}, \eqref{eq:bch-2rdmc} shows the parameters \textit{must} be identified as unitary coupled cluster amplitudes.
	\item As UDCT determines the energy by variationally minimizing an approximation to \eqref{eq:bch-energy}, it can be regarded as a VUCC method. Specifically, UDCT constructs some number of cumulant diagrams and uses a power series of their partial trace to construct an infinite sum of non-cumulant (1RDM and products of the 1RDM) terms. The mechanism of this summation is discussed in greater detail in Section \ref{sec:udct-ucc-compare}.
	\item \label{bullet:uneven} One could approximate \eqref{eq:bch-2rdmc} differently for the purposes of constructing $d$ in \eqref{eq:d-to-gamma-eval} and of constructing the $\lambda$ component of \eqref{eq:bch-2rdm}. This was done in the ODC-13 method of Reference \citenum{Sokolov:2014p074111}, which used a degree-four truncation and a degree-three truncation, respectively. Reference \citenum{Sokolov:2014p074111} attributed this truncation to different degrees for the poor performance of ODC-13. None of the other methods implemented in this work use this uneven truncation strategy.
	\item Commutator truncations are not the only way to approximate \eqref{eq:bch-2rdm} and the derived \eqref{eq:bch-2rdmc}, although they are widely used.\cite{Hoffmann:1988p993, Chen:2012p014108, Bartlett:1989p133, Watts:1989p359, Watts:1989p502} For example, there is the recursive commutator approximation,\cite{Yanai:2006p194106, Yanai:2007p104107, Neuscamman:2009p124102, Evangelista:2012p27} where high-rank second quantized operators are projected out of commutators. If only RDMs at the converged amplitudes are necessary, there are also truncations of the inherently projective Bernoulli functional.\cite{Liu:2018p244110, Hodecker:2020p094106, Hodecker:2020p084112, Hodecker:2020p3654}
\end{enumerate}

\subsection{Orbital-Optimized Unitary Methods}
\label{sec:oo}

The $t$ amplitudes appearing in \eqref{eq:def-t} through \eqref{eq:bch-2rdmc} imply a division of orbitals into occupied and virtual spaces. While most electronic structure methods relying on such a partition choose this division based on Hartree--Fock orbitals, it is possible to vary these orbitals over a computation. There are multiple possible criteria for what the converged orbitals of a computation are.\cite{Kutzelnigg:2006p171101, Nesbet:1958p1632, Purvis:1982p1910} If we perform VUCC or UDCT with the orbitals that minimize the energy, we call the resulting methods orbital-optimized variational unitary coupled cluster (OVUCC) and orbital-optimized unitary density cumulant theory (OUDCT),\cite{Sokolov:2013p204110} respectively. The stationarity conditions are functions of the reduced density matrices,\cite{Bozkaya:2011p104103} therefore DCT does not need to use \eqref{eq:gamma-lambda-deriv} to compute the derivative of the energy with respect to orbital rotations. Orbital optimized methods are well-studied,\cite{Kats:2014p061101, Sherrill:1998p4171, Bozkaya:2011p104103, Sokolov:2013p204110, Kats:2018p13, Stein:2014p214113, Robinson:2011p044113, Bozkaya:2014p204105, Pavo_evi_:2020p1578, Bozkaya:2014p2371} and orbital-optimized unitary coupled cluster has recently received attention from quantum computing.\cite{Sokolov:2020p124107, Mizukami} The impact of orbital optimization in density cumulant theory, compared to an alternative orbital convergence criterion,\cite{Kutzelnigg:2006p171101} is studied numerically in References \citenum{Copan:2014p2389} and \citenum{Sokolov:2013p204110}.

Because the orbitals are added as parameters, varying all unitary cluster amplitudes would lead to the dimension of the variational space being greater than the dimension of the total space of wavefunctions, guaranteeing a redundancy. To remedy this, the $T_1$ are set to $0$. We may qualitatively think of the $T_1$ amplitudes as corresponding to orbital rotations, because a unitary cluster operator consisting only of $T_1$ amplitudes is simply an orbital rotation.\cite{Helgaker_1988}

Adding orbital optimization to the unitary transformation of \eqref{eq:unitary} is a convenient choice for multiple reasons. First, because the exact unitary coupled cluster energy is a variational upper bound to the energy for any choice of cluster operators, the argument of K{\"o}hn and Olsen that orbital optimization costs reproducing the full configuration interaction limit does not apply.\cite{Kohn_2005} Second, eliminating the $T_1$ amplitudes reduces the number of contractions that need to be considered in \eqref{eq:bch-2rdmc}. Third, this means that in the gradient theory, it is not necessary to compute an orbital relaxation term.\cite{Sherrill:1998p4171, Sokolov:2013p204110} This both makes the analytic gradient theory simple and means there is no need to distinguish between the reduced density matrices delivered by the theory and ``relaxed density matrices'' including extra Lagrangian terms. Fourth, these operators do not need to be expanded in an infinite series in the manner of \eqref{eq:bch-2rdmc},\cite{Sokolov:2013p204110} so we may completely avoid errors due to truncation of an infinite series with these parameters.\cite{Sokolov:2014p074111, Copan:2018p4097}

The fifth reason is subtler and specific to UDCT.  Because $\gamma$ is an implicit function of $d$, which is in turn a function of the amplitudes $t$, $\gamma$ is an implicit function of the amplitudes. When the denominator of \eqref{eq:gamma-lambda-deriv} is not zero, the chain rule gives:

\begin{equation}
	\label{eq:gamma-t-deriv}
	\frac{\partial}{\partial t} \gamma^{p^\prime}_{q^\prime} = \frac{1}{\gamma_{p^\prime} + \gamma_{q^\prime} - 1} \frac{\partial}{\partial t} d^{p^\prime}_{q^\prime} \quad
\end{equation}

\noindent where $t$ is an arbitrary amplitude, and we are working in the basis of natural spin-orbitals of our current 1RDM. If orbital $p^\prime$ is occupied and $q^\prime$ is virtual, or vice versa, ${\gamma_{p^\prime} + \gamma_{q^\prime} - 1 \approx 0}$, and the denominator of \eqref{eq:gamma-lambda-deriv} becomes very small, which may produce numerical issues.

This calamity is avoidable. If $\frac{\partial}{\partial t} d^{p^\prime}_{q^\prime} = 0$, then the right side of \eqref{eq:gamma-t-deriv} is zero, even if the denominator is very close to zero. For all previously studied DCT models, this is true in the occupied-virtual blocks for any choice of $t$, so we have

\begin{equation}
	\label{eq:gamma-t-deriv-ov}
	\frac{\partial}{\partial t} \gamma^\textnormal{v}_\textnormal{o} = \frac{\partial}{\partial t} \gamma^\textnormal{o}_\textnormal{v} = 0 \quad .
\end{equation}

\noindent Let us call the orbitals used to define the amplitudes the \textit{reference orbitals}. The origin of \eqref{eq:gamma-t-deriv-ov} is that the occupied-virtual and virtual-occupied elements of $d$ are zero for any choice of $t$. This means occupied natural orbitals are linear combinations of occupied orbitals, and virtual natural orbitals are linear combinations of virtual orbitals. Combining these facts means that after moving to the current natural orbital basis for \eqref{eq:gamma-t-deriv}, $d^\textnormal{o}_\textnormal{v}$ and $d^\textnormal{v}_\textnormal{o}$ remain identically zero even as the cumulant changes. Their derivative must therefore vanish. This implies $\gamma^\textnormal{v}_\textnormal{o}$ and $\gamma^\textnormal{o}_\textnormal{v}$ vanish. Intuitively, optimizing the orbitals should account for the otherwise missing correlation in these blocks.

Unfortunately, $d^\textnormal{o}_\textnormal{v}$ and $d^\textnormal{v}_\textnormal{o}$ being zero is \textit{not} a general feature of the OUDCT ansatz. To see this, let us borrow an idea from Reference \citenum{Kutzelnigg:2006p171101} and expand $\gamma$ in \eqref{eq:rdmc-partial-trace} by $\kappa + \tau$, where $\kappa$ is the 1RDM of $\Phi$, and $\tau$ is the remainder. Using the $\Phi$-normal operator $\tilde{a}^p_q$, $\tau$ may be expressed as:
\begin{equation}
	\label{eq:bch-tau}
	\tau^{p}_{q}(t) = \sum\limits_{n = 0}^\infty \frac{1}{n!} \bra{\Phi} [\cdot, T - T^\dagger]^n(\tilde{a}^{p}_{q}) \ket{\Phi} \quad .
\end{equation}

\noindent The occupied-virtual block of both sides of \eqref{eq:rdmc-partial-trace} is given by

\begin{equation}
	\label{eq:dia}
	d^i_a = \tau^i_p \tau^p_a
\end{equation}

\noindent and we must tell when this is nonzero. If only even rank operators are included in \eqref{eq:def-big-t}, then the total excitation rank of terms in \eqref{eq:bch-tau} is odd no matter how many even rank operators are contracted against $\tilde{a}^i_a$, so no complete contractions are possible. Consequently, $\tau^i_a$ is zero by \eqref{eq:bch-tau}, and $d^i_a$ is zero by \eqref{eq:dia}.

But if all operators are present in \eqref{eq:def-big-t}, then $\tau^i_a$ terms exist and give rise to non-vanishing $d^i_a$ through \eqref{eq:dia}. There are $d^i_a$ terms of degree three in the amplitudes, but they vanish when $T_1 = 0$ due to their dependence on the $t^i_a$ term of $\tau^i_a$. Terms without $T_1$ amplitudes, and thus nonzero even with optimized orbitals, first appear at degree four due to degree two terms of $\tau^i_a$. The term that is of degree three in $T_2$ and degree one in $T_3$ is $d^i_a = (\frac{1}{8} t^{ijk}_{bcd} t^{cd}_{jk})(\frac{1}{2} t^{be}_{lm} t^{lm}_{ae}) + (-\frac{1}{2} t^{be}_{lm} t^{im}_{be})(\frac{1}{8} t^{ljk}_{acd} t^{cd}_{jk})$. While inserting the series expansions of \eqref{eq:bch-tau} into \eqref{eq:dia} can lead to cancellations, we do not observe cancellations in this example. This formula can also be derived directly, albeit tediously, from \eqref{eq:def-d}. We show this in the Supporting Information.

The implication is that continuing the OUDCT ansatz will eventually lead to the terms with small denominators in \eqref{eq:gamma-t-deriv} being multiplied by something that is not identically zero. These small denominators then must be computed, which is likely to lead to numerical problems.

We have tested our results numerically by performing exact orbital-optimized unitary coupled cluster (OUCC) on the \ce{Be} atom in the cc-pVDZ basis set. We find that the occupied-virtual block of the 1RDM, where orbital spaces are determined by the optimal reference, has a norm of $9.7 \times 10^{-5}$. This is nonzero, to machine precision. A commutator expansion of the 1RDM shows that the block is numerically zero to one commutator, but has a norm of $9.5 \times 10^{-5}$ after the second commutator. As discussed in Section \ref{sec:udct-ucc-compare}, this is the commutator at which we expect the block to first become nonzero in OUCC, and this is consistent with $d^i_a$ being nonzero at the degree four terms of the cumulant, \eqref{eq:bch-2rdmc}.

We close with a technical remark. In the special case that the occupied-virtual block of $d$ is identically zero, we can always choose the natural orbitals such that each natural orbital is obtained by diagonalizing either the occupied block or the virtual block. If we make this choice, then the 1RDM derivative, \eqref{eq:gamma-lambda-deriv}, simplifies into a series of equations for the occupied block and a series for the virtual block. This also leads to \eqref{eq:gamma-t-deriv-ov}, the occupied-virtual block of the 1RDM being zero. This construction is identical with the one from \eqref{eq:gamma-t-deriv} in the case that $\gamma_{p^\prime} + \gamma_{q^\prime} \neq 1$. But if $\gamma_{p^\prime} + \gamma_{q^\prime} = 1$, \eqref{eq:gamma-t-deriv} is not even defined, but constructing $d$ by diagonalizing the blocks of $\gamma$ separately remains valid. This block-diagonal procedure also gives a continuous $\gamma$, and if some infinitesimal change in $d$ breaks the degeneracy, this is the only choice of $\gamma$ that will be continuous in the direction of that infinitesimal change.

\subsection{Comparison of UDCT and UCC Truncations}
\label{sec:udct-ucc-compare}
Suppose that the series \eqref{eq:bch-2rdmc} is evaluated to a certain number of commutators, and the 1RDM is generated from \eqref{eq:d-to-gamma-eval}. Can we conclude that this includes all the terms from evaluating \eqref{eq:bch-1rdm} to the same number of commutators?

For a general cluster operator, we cannot. Evaluating \eqref{eq:bch-2rdmc} to one commutator gives $d = 0$, which fails to generate the one commutator contribution to $\gamma^i_a$, $t^i_a$. Evaluating \eqref{eq:bch-2rdmc} to two commutators gives a block-diagonal $d$ (by Section \ref{sec:oo}), which cannot generate any $\gamma^i_a$ terms, such as $\frac{1}{8} t_{abc}^{ijk} t_{jk}^{bc}$. Because the latter term does not contain $T_1$ amplitudes, determining $d$ to degree $n$ does not guarantee that $\gamma$ is determined to degree $n$, even for orbital-optimized methods.

To explain this puzzling fact, recall that to construct $\gamma$, \eqref{eq:soln-space} requires the change-of-basis matrix $U$ given by the eigenvectors of $d$. We may expect that to determine $\gamma$ to degree $n$ in the amplitudes, we require $U$ to degree $n$.
Unfortunately, determining $U$ to degree $n$ from \eqref{eq:def-u} requires $d$ to degree $n+2$ because $d$ is nonzero only at degree two and greater. In general, determining $d$ to degree $n+2$ suffices to determine $\gamma$ to degree $n$. In the Supporting Information, we explicitly show this is \textit{necessary} for $n = 1$ and $n = 2$, if no restrictions are put on the cluster operator. If the cluster operator is restricted so $T_1 = 0$, but no other restrictions are imposed, it is also necessary for $n = 2$.

There is an important special case where determining $\gamma$ from $d$ requires fewer commutators. If $U$ has the same block-diagonal structure as the central matrix of \eqref{eq:soln-space}, then \eqref{eq:soln-space} simplifies into a single block-diagonal matrix, both blocks of which are power series in $d$.\cite{Kutzelnigg:2006p171101} Then if $d$ is correct to degree $n$, the power series of $\gamma$ must be as well.

In this special case, which this paper will focus on, a degree $n$ truncation of OUDCT includes all the degree $n$ terms of OVUCC, plus terms of higher degree in the $t$ amplitudes. It is even possible to identify which terms of higher degree are included in this case, which we shall consider later.

\section{Computational Implementation}
\label{sec:implement}
To conduct the studies described in Section \ref{sec:numerical}, we created (a) a Python program to perform OUDCT and OVUCC computations within a given commutator truncation, (b) a Python program to perform an exact orbital-optimized unitary coupled cluster (OUCC) computation, and (c) a Python program to compare the amplitudes of a truncated OUDCT or OVUCC computation with those determined from a full OUCC computation. Because the equations for truncated unitary theories were already quite complicated,\cite{Sokolov:2014p074111} we created a code generator to derive the necessary tensor contractions for the OUDCT and OVUCC computations.

Below, we describe the code and the correctness checks we employed to ensure the accuracy of the results.\cite{code}

\subsection{Generation and Implementation of Truncated Orbital-Optimized Unitary Theories}

Our code generator first draws all possible fully closed diagrams of the normal ordered forms of \eqref{eq:unitary-1pdm} and \eqref{eq:unitary-2pdm}. Diagrams that differ only in the ordering of their operators are considered distinct. Subsequently, diagrams identical by time ordering are collected into a single expression. Diagrams of \eqref{eq:unitary-2pdm} are separated into disconnected diagrams, which are not explicitly used in OUDCT but are explicitly used in OVUCC, and the connected diagrams, which are used in both theories. For OUDCT, the connected 2RDM diagrams are partial traced to obtain $d$ of \eqref{eq:def-d} as an explicit function of the amplitudes. The residual equations are obtained by differentiation of the energy expressions. No special code is generated for the orbital optimization, as those expressions are kept in terms of the reduced density matrices.

From orbital and amplitude residuals, update steps to the orbital parameters and amplitudes were computed using a crude diagonal approximation to the exact Hessian giving denominators of signed ``orbital energies'' as in M{\o}ller--Plesset perturbation theory for OVUCC and unsigned diagonal elements of \eqref{eq:gen-fock} for OUDCT, per Section 2.2 of Reference \citenum{Wang:2016p4833}. Direct inversion of the iterative subspace (DIIS)\cite{Hamilton:1986p5728} is used to accelerate convergence of the combined vector of orbital amplitudes and $t$ amplitudes. All tensor contractions use the \texttt{opt\_einsum} package for efficiency,\cite{G_A_Smith:2018p753} and all integrals are obtained from a developer version of \textsc{Psi4} 1.4.\cite{Smith:2020p184108, Smith:2018p3504} To ensure tight convergence, we required that the norms of the amplitude gradient and orbital gradient were both under $1 \times 10^{-12}$. To address convergence problems discussed in Section \ref{sec:h2}, we enabled reading in amplitudes and overlap-corrected molecular orbital coefficients from previous computations.

To confirm the accuracy of the generated equations, we performed various checks. To confirm the accuracy of our expressions for the reduced density matrices, we compared our degree-four expressions for \eqref{eq:bch-2rdmc} to those previously published.\cite{Sokolov:2014p074111} We also implemented a code for \textit{projective} unitary coupled cluster with Hartree--Fock orbitals and confirmed that our energies match the previously reported energies for the commutator truncations studied. \cite{Evangelista:2011p224102} Although the conditions for determining amplitudes differ for projective and variational unitary coupled cluster, the function from amplitudes to energy is the same. Hence, by confirming the correctness of the functional for the projective case, we have confirmed its correctness in the variational case.

To confirm the accuracy of our expressions for the $d$ matrices, we performed OUDCT computations both with our explicit expression and by simply taking the partial trace of \eqref{eq:def-d}. In both cases, we observed the same energy. To confirm the accuracy of our computed derivatives, we have computed the dipoles for all OVUCC and OUDCT truncations studied both by finite difference and analytically, since both OVUCC and OUDCT automatically deliver relaxed density matrices suitable for property computations, by the Hellmann--Feynmann theorem. If the orbitals or amplitudes do not variationally minimize the energy function, the Hellmann--Feynmann theorem does not apply, and the two dipoles will differ. In all cases, we found that the two matched to ten or more decimal places. Consequently, we have also been able to implement the analytic gradients of these theories.

Excluding ODC-13, all OUDCT methods studied use the same parameterization of the cumulant for the intermediate $d$ \eqref{eq:def-d} as for reconstructing the RDM \eqref{eq:2rdm-decomp}. As a consequence, the partial trace satisfies the equation $\gamma^{pr}_{qr} = \gamma^p_q (\gamma^r_r - 1)$. We have numerically confirmed this for all OUDCT truncations. We note that OUDCT truncations do not necessarily satisfy \eqref{eq:rdm-partial-trace} or \eqref{eq:rdm-partial-trace2}, where $n$ is the integer number of electrons, but the OVUCC truncations do. This is because \eqref{eq:rdm-partial-trace} and \eqref{eq:rdm-partial-trace2} are true as polynomials in the unitary coupled cluster amplitudes appearing through \eqref{eq:unitary-1pdm} and \eqref{eq:unitary-2pdm}, so the partial trace of each degree in $t$ must be zero. OVUCC either includes all or none of the terms of a given degree, but OUDCT does not, as described in Section \ref{sec:udct-ucc-compare}.

\subsection{Exact Orbital-Optimized Unitary Coupled Cluster}

We implemented a scheme to obtain the exact OUCC orbitals and amplitudes via projective unitary coupled cluster. (When the cluster operator is not truncated, variational and projective unitary coupled cluster are equivalent. When only $T_1$ is removed, the same is true of their orbital-optimized variants.) Our algorithm consists of macroiterations and microiterations.

In each macroiteration, we solve the projective UCC equations exactly through the microiterations. This gives us a wavefunction of form $\exp(T-T^\dagger) \ket{\Phi}$. If the norm of $T_1$ is less than $1 \times 10^{-8}$, we have converged to the exact amplitudes and orbitals, and the algorithm terminates. Otherwise, we make a guess to the exact orbitals as $\exp(T_1 - T_1^\dagger) \ket{\Phi}$ and proceed to the next macroiteration.

We solve the projective UCC equations in a given one-electron basis set following the prescription of Evangelista.\cite{Evangelista:2011p224102} We construct the Hamiltonian, $H$, and $T - T^\dagger$ in the basis of determinants. We then compute $\exp(T^\dagger - T) H \exp(T-T^\dagger) \ket{\Phi}$, where $\Phi$ is the reference determinant, via the built-in matrix exponential and matrix multiplication operators of NumPy.\cite{Harris2020array} If $\exp(T-T^\dagger)\ket{\Phi}$ is an exact eigenstate, then

\begin{equation}
	\label{eq:pucc}
	\exp(T^\dagger - T) H \exp(T-T^\dagger) \ket{\Phi} = E \ket{\Phi}
\end{equation}

\noindent so we select amplitudes such that the projection of $\exp(T^\dagger - T) H \exp(T-T^\dagger) \ket{\Phi}$ onto all excited determinants is zero. We take steps according to the formula

\begin{equation}
	\Delta t^I_A = \frac{- r^I_A}{\varepsilon^I_A \phi^I_A}
\end{equation}

\noindent where $I$ is equivalent to the occupation vector of the orbitals excited from, $A$ is equivalent to the occupation vector of the orbitals excited to, $\varepsilon$ is the sum of virtual orbital energies minus occupied orbtial energies, and $\phi^I_A$ is the phase factor between the reference determinant and the relevant excited determinant. Convergence of the amplitudes within a given macroiteration is accelerated by DIIS.\cite{Hamilton:1986p5728} In all cases, we enforce convergence when the difference of two sides of \eqref{eq:pucc} has norm less than $1 \times 10^{-9}$. We find that by this point, the energy is converged to within $1 \times 10^{-14}$ Hartrees. As a final correctness check, the energy is then compared against the full configuration interaction energy from \textsc{Psi4}.

\subsection{Amplitude Comparisons}
When comparing amplitudes from exact OUCC and an approximate OUCC, the two amplitudes of the two methods are \textit{not} in the same one-electron basis. To compare these quantities, after constructing them, we move all quantities to the basis of the approximate theory. We then find the difference of the two quantities and compute its Euclidean norm, the square root of the sum of squares of the elements. In the context of matrices, this has also been called the Frobenius norm. This metric is invariant to unitary choices of basis to perform the comparison in, and couples how well the amplitudes match with how well the orbitals match.

\section{Results and Discussion}
\label{sec:numerical}

In this section, we consider the OUDCT and OVUCC ans{\"a}tze truncated to $T = T_2$, with between two to five commutators. We call these methods by names such as OUDCT$n$ and OVUCC$n$, where $n$ denotes the number of commutators. OUDCT2 is also known as ODC-12,\cite{Sokolov:2013p204110} and OVUCC2 is also known as OCEPA(0).\cite{Bozkaya:2013p054104} We shall also consider ODC-13,\cite{Sokolov:2014p074111} which cannot be expressed as a single commutator truncation.

\subsection{\ce{H2} Dissociation}
\label{sec:h2}
The dissociation of \ce{H2} has been previously used to model the performance of DCT over a range of electron correlation strengths.\cite{Sokolov:2013p204110, Sokolov:2014p074111, Copan:2018p4097} OUDCT with only a $T_2$ operator will be exactly OVUCC with only a $T_2$ operator in the limit of an infinite commutator expansion. Because \ce{H2} is a two-electron system, if UCC is exact, OVUCC with only a $T_2$ operator will be exact as well. Due to this close relationship between the OVUCC and OUDCT ans{\"a}tze, we first compute the dissociation curve with the truncated OVUCC ansatz for comparison.

\subsubsection{Potential Energy Curves}
Previous experience with commutator truncations of projective UCC with Hartree--Fock orbitals suggests that the stronger correlation effects are, the less accurate a given UCC commutator truncation should be.\cite{Evangelista:2011p224102}

\begin{figure}
	\caption{Dissociation curves of \ce{H2} from 0.6 \AA\ to 2.5 \AA\ computed with low-degree commutator truncations of the (\protect\subref{subfig:H2-energy-ucc}) OVUCC and (\protect\subref{subfig:H2-energy-dct}) OUDCT ans{\"a}tze in the cc-pVDZ basis set.}
	\label{fig:H2-energy}
	\begin{subfigure}{0.45\textwidth}
		\caption{}
		\label{subfig:H2-energy-ucc}
		\includegraphics[width=\linewidth]{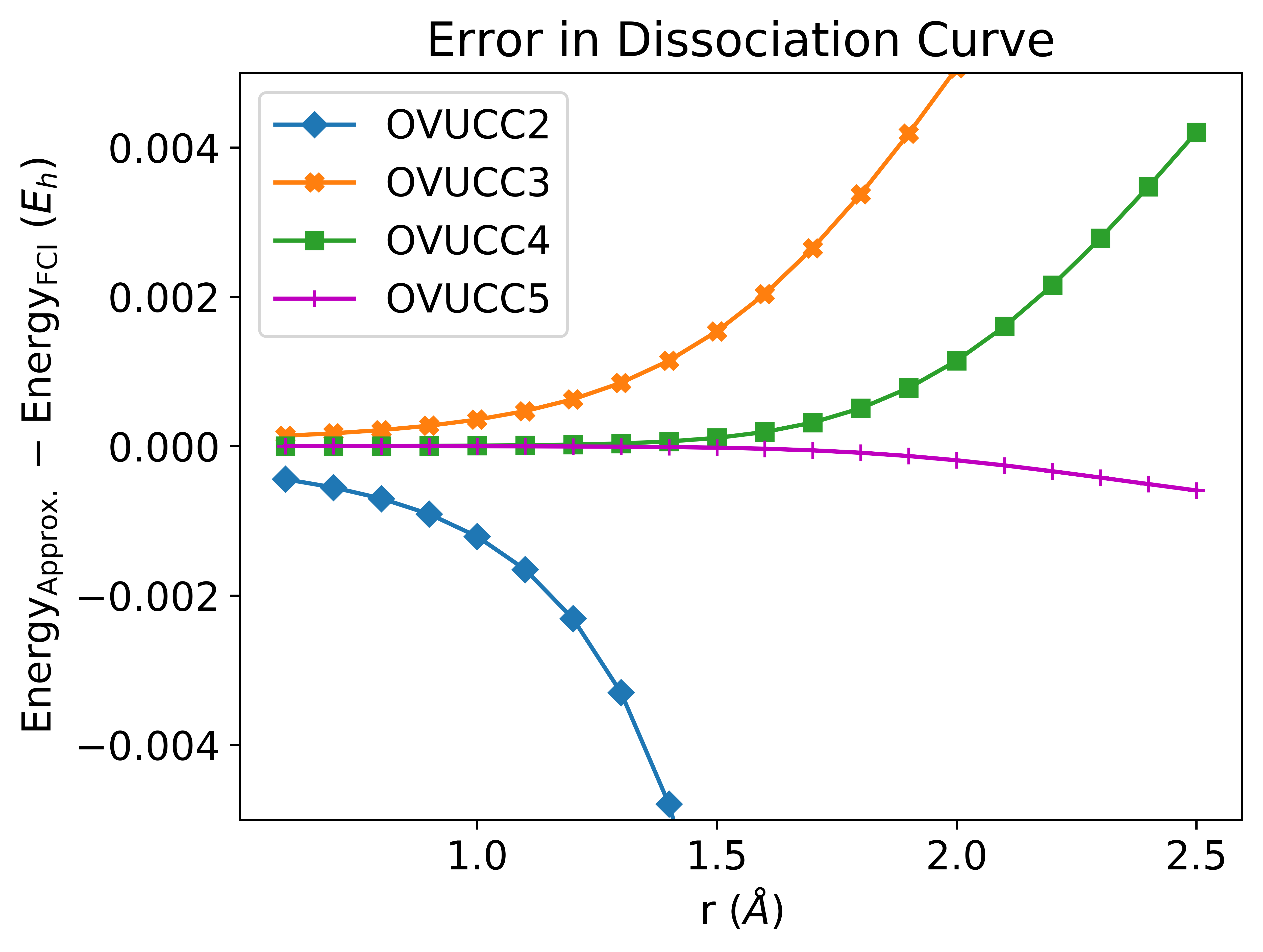}
	\end{subfigure}
	\begin{subfigure}{0.45\textwidth}
		\caption{}
		\label{subfig:H2-energy-dct}
		\includegraphics[width=\linewidth]{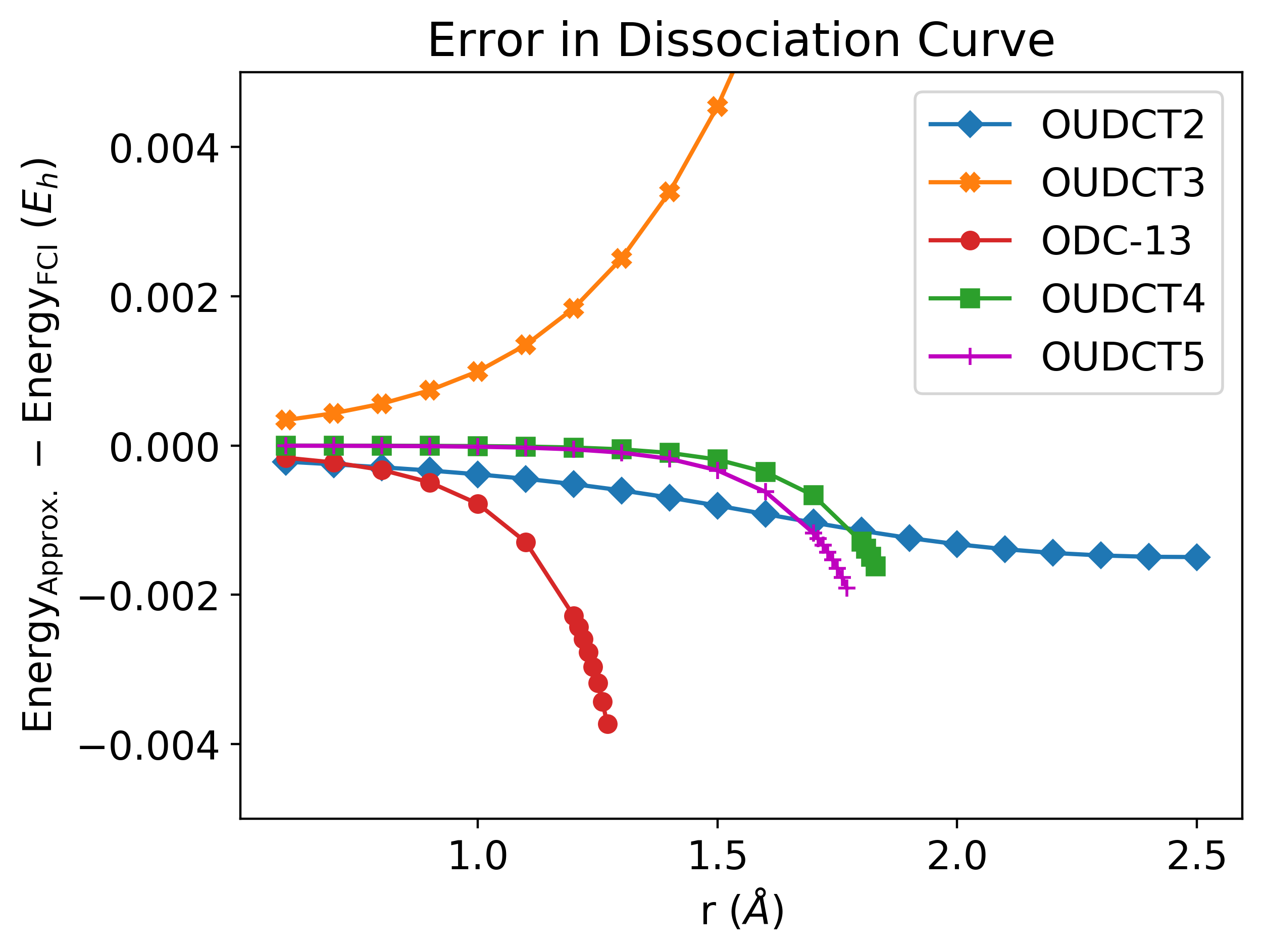}
	\end{subfigure}
\end{figure}

OVUCC illustrates this trend, as well as smooth convergence with respect to the number of commutators. The error in the energy curve for \ce{H2} is shown in Figure \ref{subfig:H2-energy-ucc}. For every geometry, adding another commutator decreases the error of the energy compared to FCI. When going from OVUCC2 to OVUCC3, or OVUCC4 to OVUCC5, the decrease is roughly by a factor of five. When going from OVUCC3 to OVUCC4, the decrease is by a factor of 100 in the equilibrium region, but diminishes to about a factor of two by around 2.5 \AA. That odd and even rank truncations of VUCC will perform differently was theorized by Kutzelnigg.\cite{Kutzelnigg:1982p3081} We observe no convergence problems except for OVUCC2, also known as OCEPA(0).\cite{Bozkaya:2013p054104} The poor performance of OCEPA(0) for \ce{H2} dissociation has been reported previously\cite{Sokolov:2013p204110} and is unsurprising, as singularities are known to appear in CEPA(0) for bond dissociation.\cite{Taube:2009p144112}

OUDCT displays markedly different behavior across two regimes, shown in Figure \ref{subfig:H2-energy-dct}. For near-equilibrium geometries, with the exception of OUDCT3, we observe improved accuracy as more commutators are added, as shown by the equilibrium geometry and harmonic vibrational frequency in Table \eqref{tbl:h2-eqn}.

\begin{table*}
	\caption{ Errors in the equilibrium bond length and harmonic vibrational frequencies of \ce{H2}, relative to FCI, for approximate OUDCT and OVUCC methods with $T = T_2$, using the cc-pVDZ basis set.
		\label{tbl:h2-eqn}}
	\begin{tabular*}{15.5cm}{c|ccccccccc}
			 & OUDCT2 & OUDCT3 &ODC-13  & OUDCT4 & OUDCT5 &  OVUCC2 & OVUCC3 & OVUCC4 & OVUCC5 \\
		 & (ODC-12) & & & & & (OCEPA(0)) & & & \\ \hline
		$r_e$ (pm) & 0.03 & --0.10 & 0.08 & 0.00 & 0.00 & 0.12 & --0.03 & 0.00 & 0.00 \\
		$\omega$ (\cm)  & --5 & 21 & --20 & --1 & --1 & --25 & 12 & 0 & 1\\
	\end{tabular*}
\end{table*}

But as the bond stretches, the various OUDCT methods behave dramatically differently. In agreement with previous studies,\cite{Sokolov:2013p204110} OUDCT2, also known as ODC-12, has robust performance. In contrast to the other models, its error curve does \textit{not} have an exponential shape. OUDCT3 can be converged, but the energy error increases sharply, and is on the same order of magnitude as the energy errors of OVUCC2. OUDCT4 can only be converged with difficulty for internuclear distances greater than 1.6 \AA. We were not able to converge the equations after 1.8 \AA, even reading in the amplitudes and orbitals (after accounting for the change in overlap matrix) from previous computations. OUDCT5 is similar, but with larger energy errors.

ODC-13 also encounters significant error before diverging around 1.3 \AA, which was attributed to large violations in the equations \eqref{eq:rdm-partial-trace2} and $\gamma^{pq}_{pq} = n (n-1)$, which hold for pure $n$-representable RDMs.\cite{Sokolov:2014p074111} OUDCT4 and OUDCT5 follow these equations more closely than the OUDCT2 method by an order of magnitude. See Supporting Information for the errors in \eqref{eq:rdm-partial-trace2}. While we observe that the partial trace failure of ODC-13 contributes to its poor performance, using more consistent approximations only delays the convergence problems.

\subsubsection{Amplitude Residuals}
\label{sec:residuals}
\begin{figure}
	\caption{The difference between the converged doubles amplitudes for (\protect\subref{subfig:H2-RN-ucc}) OVUCC and (\protect\subref{subfig:H2-RN-dct}) OUDCT theories and the exact unitary $T_2$ amplitudes as a fraction of the norm of the exact OVUCC amplitudes for \ce{H2} computed with the cc-pVDZ basis set.}
	\label{fig:H2-R}
	\begin{subfigure}[b]{0.45\textwidth}
		\caption{}
		\label{subfig:H2-RN-ucc}
	\includegraphics[width=\linewidth]{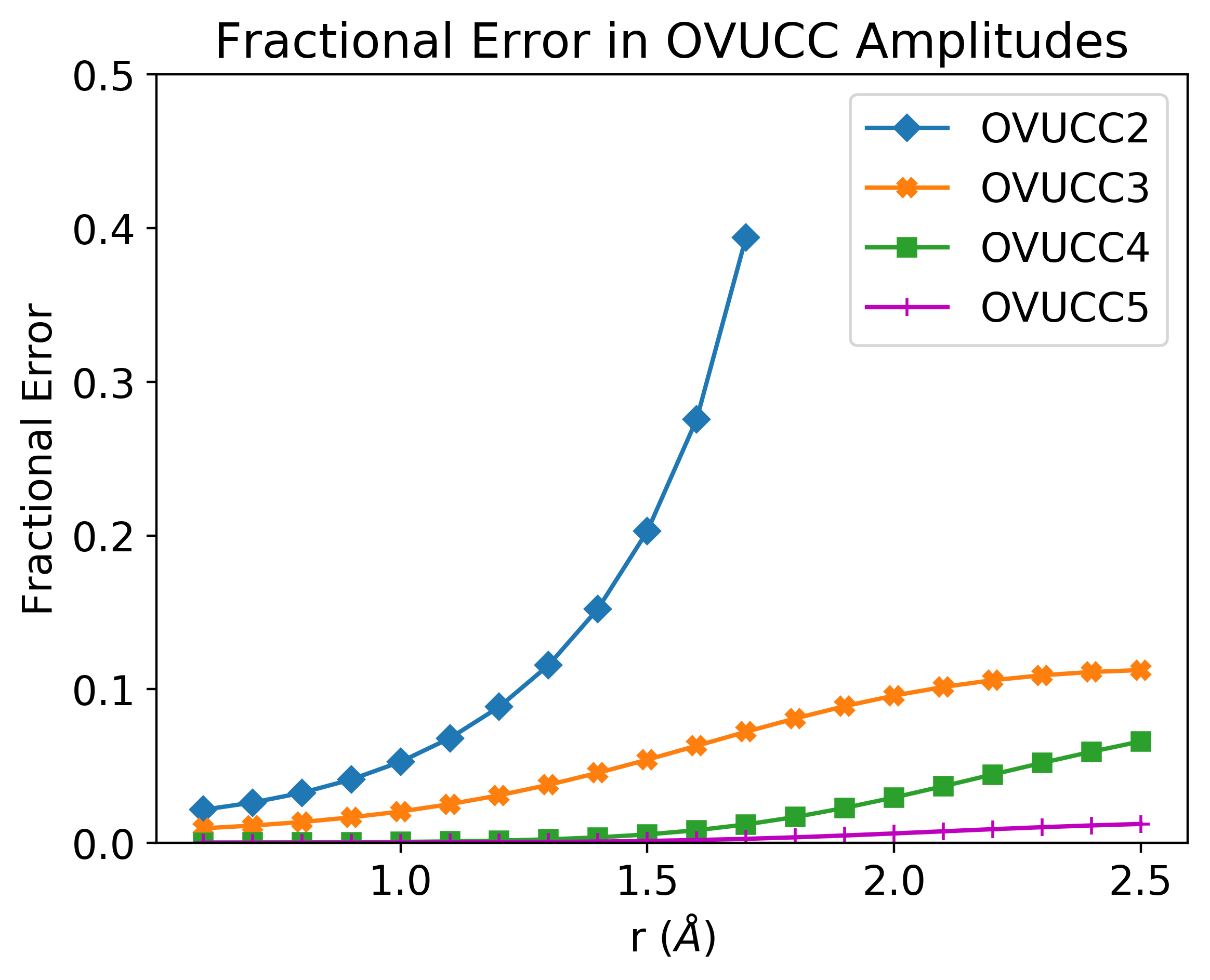}
	\end{subfigure}
	\begin{subfigure}[b]{0.45\textwidth}
		\caption{}
		\label{subfig:H2-RN-dct}
	\includegraphics[width=\linewidth]{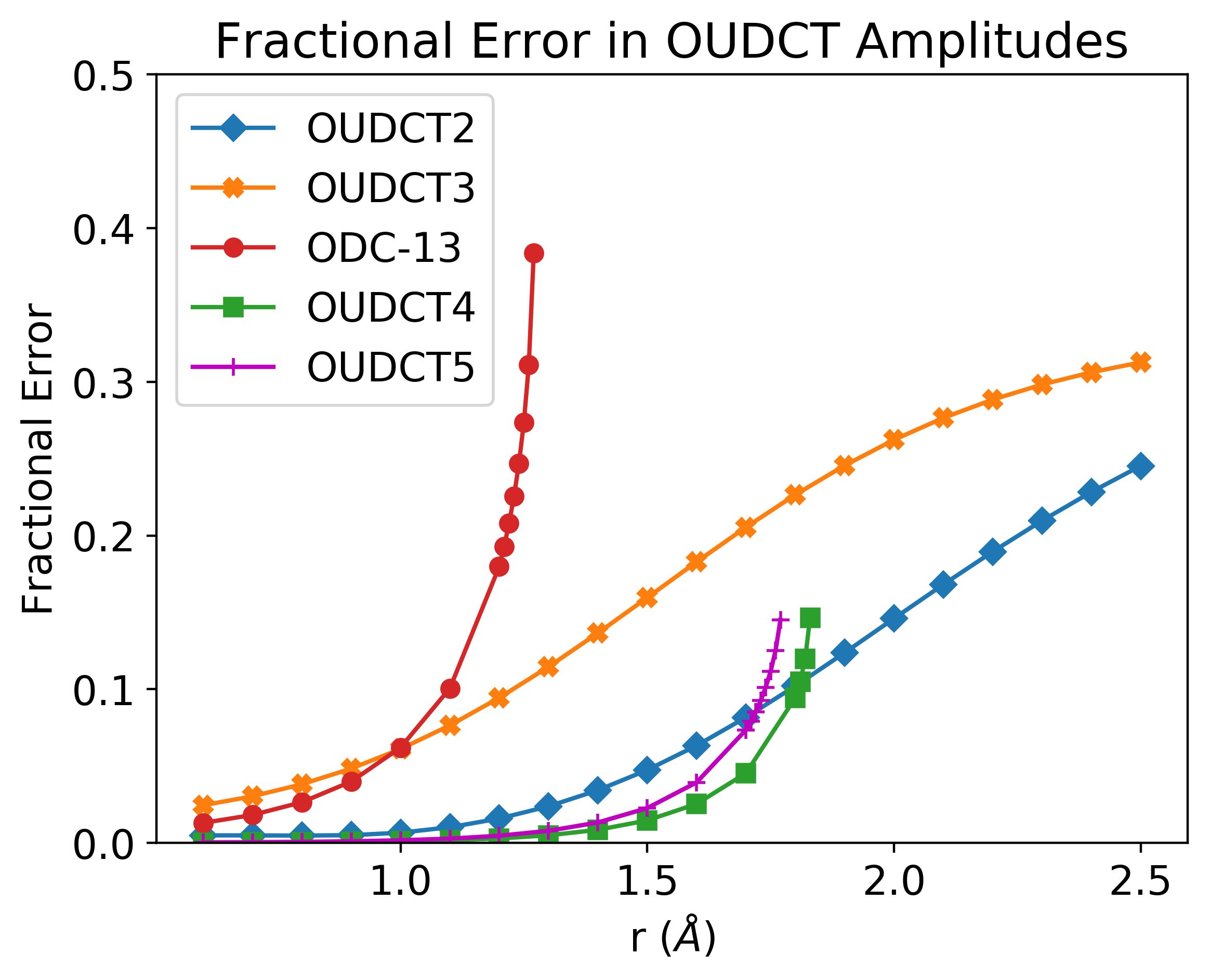}
	\end{subfigure}
\end{figure}

It may seem puzzling that OUDCT2, also known as ODC-12, yields a relatively accurate \ce{H2} dissociation curve, but less severe truncations of the same ansatz lead to more severe errors in dissociation curves. Intuition would suggest that OVUCC2 is already a good approximation to the exact OUDCT ansatz, and better approximations to the ansatz would give better energies. To identify the flaw in this inuition, consider the difference between the exact $t$ amplitudes and the final $t$ amplitudes of the approximate computation as a fraction of the norm of the exact amplitudes, shown in Figure \ref{subfig:H2-RN-ucc} for OVUCC and  Figure \ref{subfig:H2-RN-dct} OUDCT.

For all OVUCC and most OUDCT truncations, low error in the energies coexist with low error in the amplitudes, compared to the exact OUCC theory. For OVUCC, less severe commutator truncations decrease both errors across the entire curve, but for OUDCT, this decrease only occurs at weakly correlated geometries. For stretched geometries, OUDCT's error is much worse. We must attribute this to OUDCT's partial inclusion of terms of degree higher than the truncation level, as discussed in Section \ref{sec:udct-ucc-compare}. At weakly correlated geometries, the small value of the amplitudes means these high degree terms are negligible, leading to good accuracy.

Comparing Figures \ref{subfig:H2-RN-ucc} and \ref{subfig:H2-RN-dct}, one would expect an energy error curve for OUDCT2 an order of magnitude worse than OVUCC3, let alone OVUCC4 or OVUCC5. This is not the case. In OUDCT2, an accurate \ce{H2} dissociation curve coexists with large errors in the amplitudes, indicating that the static correlation tolerance of OUDCT2 does not result from well-approximating the exact OVUCC ansatz, at least for \ce{H2}. This suggests the effect of better approximating the exact OVUCC ansatz will be difficult to predict, in agreement with Section \ref{sec:h2}.

Alternative qualitative explanations for OUDCT2's static correlation tolerance are beyond the scope of this work, but we intend to explore this in future.

\subsection{Equilibrium Properties of Diatomics}
\label{sec:dynamic}
\begin{table*}
	\caption{ Errors in the geometries of diatomics (pm), relative to CCSDTQ(P), for approximate OUDCT and OVUCC methods with $T = T_2$, using the cc-pCVDZ basis set. $\Delta\textsubscript{abs}$ denotes the mean absolute error, and $\Delta\textsubscript{std}$ denotes the standard deviation of signed errors.
	\label{tbl:dynamic-geom}}
	\begin{tabular*}{15.5cm}{c|ccccccccc}
		Molecule & OUDCT2 & OUDCT3 &ODC-13  & OUDCT4 & OUDCT5 &  OVUCC2 & OVUCC3 & OVUCC4 & OVUCC5 \\
		& (ODC-12) & & & & & (OCEPA(0)) & & & \\ \hline
		\ce{N2} & --0.51 & --1.00 & --0.60 & --0.59 & --0.61 & --0.26 & --0.90 & --0.60 & --0.62\\
		\ce{CO} & --0.63 & --0.92	& --0.68 &  --0.66 & --0.68 & --0.50 & --0.84	& --0.66 & --0.68\\
		\ce{N2+} & --0.69 & --1.36 & --0.73 & --0.76 & --0.82 & --0.14 & --1.22 & --0.78 & --0.83\\
		\ce{BO} & --0.84 & --1.10	& --0.88 &--0.84 & --0.86 & --0.68 & --1.01 & --0.84 & --0.86\\
		\ce{CN} & --0.69 & --1.45	& --0.84 & --0.87 & --0.91 & --0.23 & --1.30 & --0.88 & --0.92\\
		\ce{NF} & --0.65 & --1.41	& --0.95 & --1.02 & --1.02 & --0.37 & --1.29 & --1.03 & --1.03\\
		\ce{NO} & --0.73 & --1.32	& --0.92 & --0.89 & --0.91 & --0.47 & --1.21 & --0.90 & --0.92\\
		\ce{BeO} & --1.56 & --1.81 & --1.50 & --1.39 & --1.40 & --1.31 & --1.65 & --1.39 & --1.44\\ \hline
		$\Delta\textsubscript{abs}$ & 0.79 & 1.30 & 0.89 & 0.88 & 0.91 &-0.50 & 1.18 & 0.89 & 0.91\\
		$\Delta\textsubscript{std}$ & 0.33 & 0.29 & 0.27 & 0.25 & 0.25 & 0.37 & 0.26 & 0.25 & 0.25\\
	\end{tabular*}
\end{table*}

\begin{table*}
	\caption{ Errors in the harmonic frequencies of diatomics (\cm), relative to CCSDTQ(P), for approximate OUDCT and OVUCC methods with $T = T_2$, using the cc-pCVDZ basis set. $\Delta\textsubscript{abs}$ denotes the mean absolute error, and $\Delta\textsubscript{std}$ denotes the standard deviation of signed errors.
		\label{tbl:dynamic-freq}}
	\begin{tabular*}{15.5cm}{c|ccccccccc}
		Molecule & OUDCT2 & OUDCT3 &ODC-13  & OUDCT4 & OUDCT5 &  OVUCC2 & OVUCC3 & OVUCC4 & OVUCC5 \\
		& (ODC-12) & & & & & (OCEPA(0)) & & & \\ \hline
		\ce{N2} & 56 & 116 & 67  & 66 & 69 & 21 & 105 & 67 & 70\\
		\ce{CO} & 67 & 91 & 72 & 68 & 70 & 55 & 85 & 68 & 70 \\
		\ce{N2+} & 50 & 137 & 54 & 73 & 79 & --38 & 124 & 75 & 81\\
		\ce{BO} & 69 & 88 & 72 & 68 & 70 & 56 & 82 & 69 & 70\\
		\ce{CN} & 46 & 106 & 56 & 59 & 63 & --1 & 95 & 60 & 64\\
		\ce{NF} & 20 & 66 & 40 & 47 & 47 & 3 & 61 & 48 & 47\\
		\ce{NO} & 71 & 132 & 92 & 89 & 92 & 43 & 122 & 90 & 93\\
		\ce{BeO} & 55 & 69 & 53 & 47 & 50 & 40 & 61 & 47 & 50 \\ \hline
		$\Delta\textsubscript{abs}$ & 54 & 101 & 63 & 65 & 68 & 32 & 92 & 66 & 68\\
		$\Delta\textsubscript{std}$ & 16 & 25 & 15 & 13 & 14 & 31 & 23 & 13 & 14\\
	\end{tabular*}
\end{table*}

To assess the performance of OUDCT methods for equilibrium properties of diatomics in systems of more than two electrons, we have computed the equilibrium geometries and harmonic vibrational frequencies of eight diatomics using OUDCT and OVUCC methods with the cluster operator \eqref{eq:def-big-t} truncated to $T_2$ and the cc-pCVDZ basis set. To exclude non-Born--Oppenheimer effects and basis set convergence, we compare to high level \textit{ab initio} results. Specifically, we compare to properties computed at the CCSDTQ(P)/cc-pCVDZ level with MRCC,\cite{K_llay:2005p214105, K_llay:2020p074107} driven using \textsc{Psi4}.\cite{Smith:2020p184108} The necessary gradients were computed by finite difference of energies. Frequencies for these systems were computed using the DIATOMIC module of \textsc{Psi4}. For the OUDCT and OVUCC methods, we instead computed gradients analytically and frequencies by finite difference of gradients with a five-point stencil.

\begin{figure}
	\caption{The mean absolute error and standard deviation of the signed errors in the (\protect\subref{subfig:dynamic-geom}) geometries and (\protect\subref{subfig:dynamic-freq}) frequencies of diatomics, relative to CCSDTQ(P), for approximate OUDCT and OVUCC methods with $T = T_2$, using the cc-pCVDZ basis set.}
	\label{fig:dynamic-graphics}
	\begin{subfigure}[b]{0.45\textwidth}
		\caption{}
		\label{subfig:dynamic-geom}
		\includegraphics[width=\linewidth]{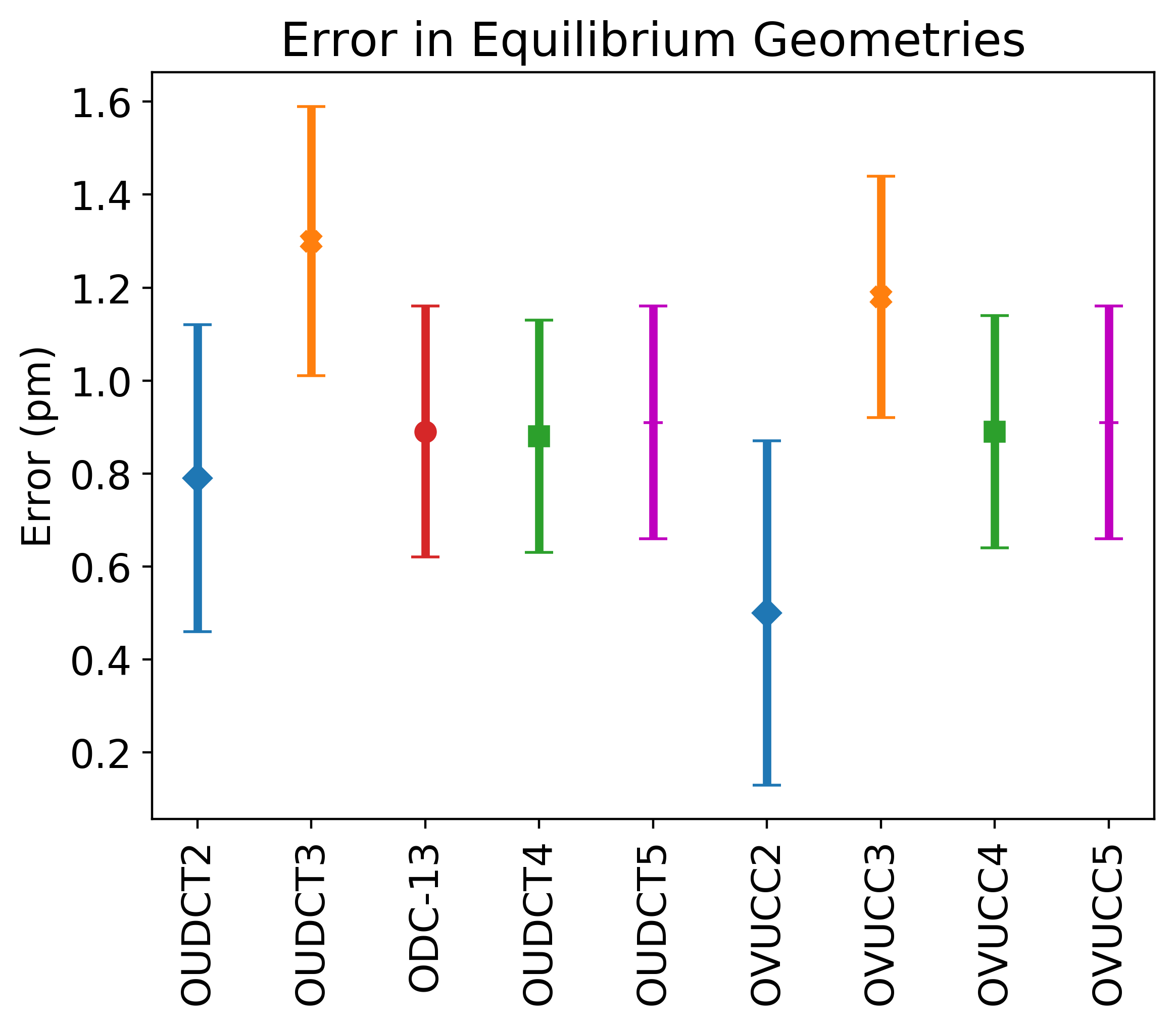}
	\end{subfigure}
	\begin{subfigure}[b]{0.45\textwidth}
		\caption{}
		\label{subfig:dynamic-freq}
		\includegraphics[width=\linewidth]{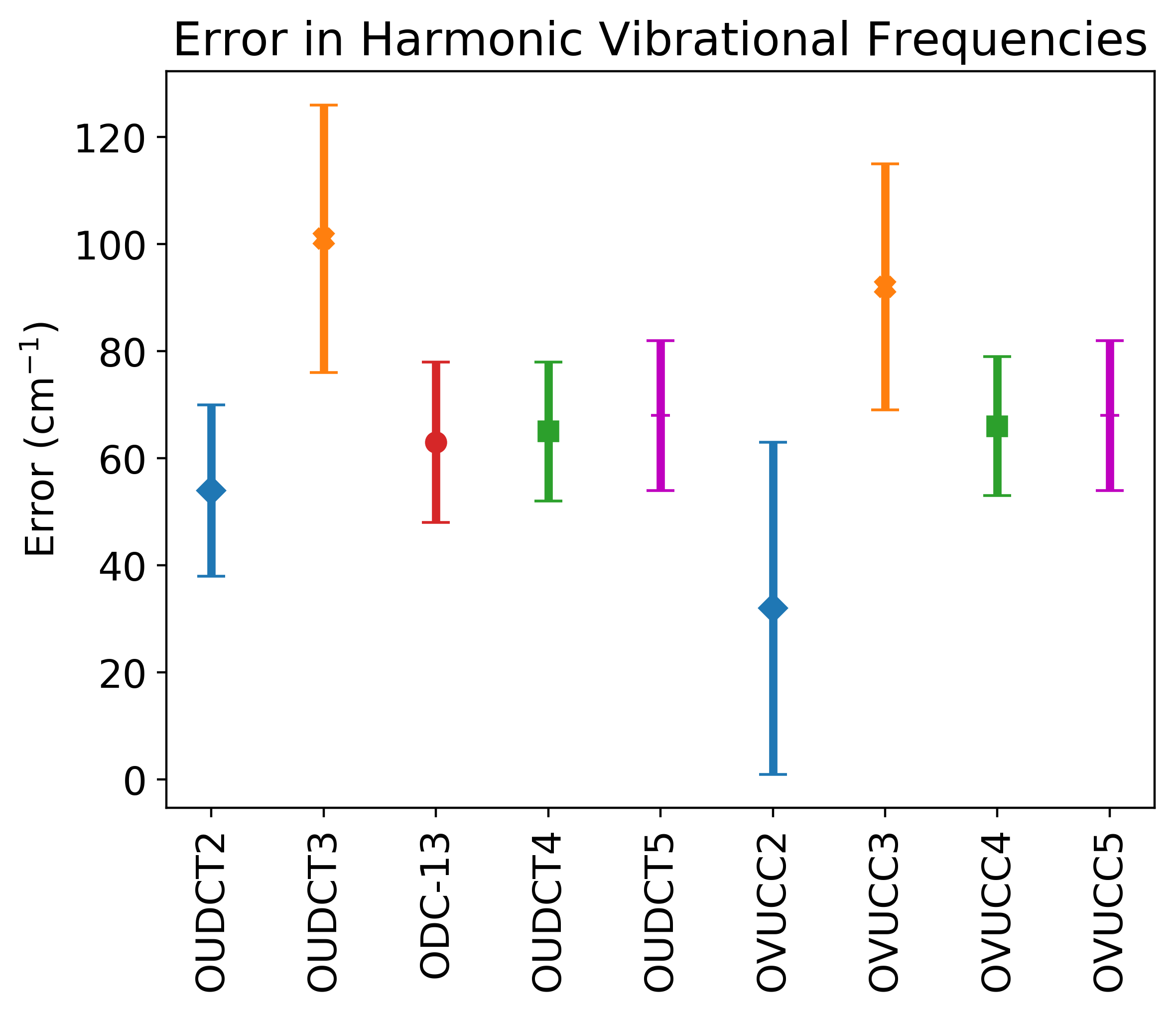}
	\end{subfigure}
\end{figure}

The errors in the equilibrium geometries are shown in Figure \ref{fig:dynamic-graphics}, and individual data points for equilibrium geometries and harmonic frequencies are given in Tables \ref{tbl:dynamic-geom} and \ref{tbl:dynamic-freq}, respectively. The ordering of methods in terms of accuracy is consistent across both sets of benchmark data. The best performance is displayed by OVUCC2, more commonly known as OCEPA(0).\cite{Bozkaya:2013p054104} The mean unsigned error in bond lengths is 0.50 pm, and the same in harmonic vibrational frequencies is 32 \cm. Second best is OUDCT2, or ODC-12,\cite{Sokolov:2013p204110} with respective mean unsigned errors of 0.79 pm and 54 \cm.

All higher order methods display slightly worse performance, on average. The degree three truncations do exceptionally poorly. The four and five commutator truncations have extremely similar performance. Mean signed geometry errors range from 0.88 to 0.91 pm, and mean unsigned harmonic frequency errors range from 65 to 68 \cm. The ODC-13 method is also similar with 0.89 pm and 63 \cm\ errors, respectively.

We reach two conclusions from this data. First, the performance of ODC-13 for equilibrium properties of diatomic molecules is unrelated to point \ref{bullet:uneven} of Section \ref{sec:udct}. The performance difference between the theory where the degree four connected terms of \eqref{eq:bch-2rdm} are included (OUDCT4) and the theory where they are not (ODC-13) is statistically insignificant.

Second, the similarity of these results upon increasing commutator truncations suggest that by four commutators, the equilibrium properties of these systems are well-converged to the exact result with respect to the number of commutators, and the difference between OVUCC and OUDCT is negligible. This is supported by our findings that both energy and amplitudes are well-converged by four or five commutators for \ce{H2} near equilibrium in Section \ref{sec:h2}. To outperform OUDCT2, OUDCT5 would need to lower $\Delta\textsubscript{abs}$ by 12 pm and 13 \cm\ for equilibrium bond lengths and harmonic frequencies, respectively. Accordingly, we expect that not even the exact OVUCC and OUDCT doubles theories (they are identical) can out-perform OVUCC2 or OUDCT2. This strongly suggests that to improve beyond OVUCC2 and OUDCT2 within orbital-optimized unitary ans{\"a}tze, it will be necessary to consider cluster operators beyond doubles. To our knowledge, the only studies of unitary cluster operators beyond doubles are the recent work of Li and Evangelista,\cite{Li:2020p234116} focused on their driven similarity renormalization group, and the non-iterative $\lambda_3$ correction considered within density cumulant theory.\cite{Sokolov:2014p074111}

\section{Conclusions}

In this research, we have studied the orbital-optimized unitary ansatz for density cumulant theory (OUDCT) both formally, and with numerical simulations of \ce{H2} dissociation and the equilibrium geometries and frequencies of diatomic molecules using low order truncations of the OUDCT ansatz with a cluster operator truncated to double excitations and de-excitations. We have also performed these simulations on analagous truncations of the closely related orbital-optimized variational unitary coupled cluster (OVUCC) ansatz. We find that:
\begin{enumerate}
	\item The DCT ansatz will encounter near-zero denominators in the gradient of the energy with respect to amplitudes if the occupied-virtual and virtual-occupied blocks of the 1-electron reduced density matrix (1RDM) are not identically zero. The OUDCT ansatz does \textit{not} preserve this property once odd-rank cluster operators are added to the ansatz. The terms that cause these problems will first appear at degree four in the Baker--Campbell--Hausdorff expansion of the density cumulant, \eqref{eq:bch-2rdmc}. If $T_1$ is included in the cluster operator, they first appear at degree three.
	\item The relationship between the OUDCT ansatz and the OVUCC ansatz is complicated by the presence of nonzero occupied-virtual and virtual-occupied blocks of the 1RDM. If these blocks are identically zero, OUDCT truncated to $n$ commutators is OVUCC truncated to $n$ commutators plus 1RDM and disconnected 2RDM terms of degree greater than $n$ in the amplitudes. If these blocks are not identically zero, OUDCT truncated to $n$ commutators will have all 1RDM terms truncated to $n-2$ commutators, but may miss terms at commutators $n-1$ and $n$.
	\item Making less severe truncations of the OUDCT ansatz does not \textit{uniformly} improve the description of the \ce{H2} dissociation curve. While it is strongly improved near equilbrium, the degree four and five theories show worse performance and convergence problems not present for the simple two-commutator truncation, ODC-12, away from equilibrium. The same is not true for OVUCC, where the same truncation procedure improves the entire curve.
	\item Making less severe truncations of the OUDCT ansatz with doubles does not improve the description of the equilibrium properties of diatomics. Including the terms from three, four, and five commutators from OUDCT and OVUCC tends to cause a minor loss of accuracy compared to the two-commutator truncations, ODC-12 and OCEPA(0). Based on the rate of convergence with respect to commutator truncation, even the doubles-only OVUCC theory with no commutator truncation is likely inferior to ODC-12 and OCEPA(0).
\end{enumerate}

Let us remark on what these results mean for future developments of DCT. If one decides to develop the theory via the OUDCT ansatz, then to improve the description of molecules at equilibrium, the results of Section \ref{sec:dynamic} advise against better approximating OVUCC doubles, and in favor of including effects of higher rank cluster operators. Triples seem to be especially important in unitary theories, as in tradiational coupled cluster theory,\cite{Li:2020p234116} and the non-iterative $\lambda_3$ correction was seen to improve DCT results.\cite{Sokolov:2014p074111} However, two special dangers then arise:

\begin{enumerate}
	\item Triples approximations must avoid near-zero denominators in \eqref{eq:gen-fock}. We see three ways to control these singularities. First, the choice of energy-minimizing orbitals while neglecting $T_1$ could be replaced in favor of the natural orbitals, where the occupied-virtual block of the 1RDM is identically zero. Second, choose the amplitudes non-variationally so that \eqref{eq:gen-fock} and its singularities are irrelevant. Either of these options makes a parameter non-variational and results in a more complicated and expensive analytic gradient theory. Third, refuse to consider theories where the block-diagonal structure of the 1RDM is compromised during iterations.  Such an approach cannot account for the nonzero terms in the occupied-virtual block of the 1RDM and cannot converge to the exact theory.  Done perturbatively, the success of the $\lambda_3$ correction\cite{Sokolov:2014p074111} suggests that this route can still be quite accurate. For iterative approaches, the first terms in the BCH expansion that must be neglected are of degree four and have both doubles and triples. Neglecting this means that degree two terms in the 1RDM are also neglected, as shown in Section \ref{sec:udct-ucc-compare}. Nonetheless, our unpublished numerical results indicate that including iterative triples to the degree two truncation of the cumulant, \eqref{eq:bch-2rdmc}, can still be quite accurate near equilibrium.
	\item The cumulant parameterization determines the accuracy of the DCT theory, but the relationship between the degree of truncation of the OUDCT ansatz and the accuracy of the resulting theory is not straightforward. Section \ref{sec:h2} demonstrates that including more commutators in the OUDCT ansatz can make the numerical results significantly less accurate. We expect the relationship to be even more complicated once triples amplitudes are included.
\end{enumerate}

However, if one wishes to improve static correlation tolerance, then the proper description of \ce{H2} is a prerequisite, and this cannot be accomplished by adding triples. Section \ref{sec:residuals} makes clear that the static correlation tolerance of ODC-12 does not originate in the method well-approximating the OVUCC ansatz. If one wishes to make a multireference generalization of ODC-12, there is not an obvious feature of ODC-12 that causes its static correlation tolerance and is therefore worth generalizing.

Further theoretical developments in DCT are needed to provide a path to more and more accurate methods. We intend to investigate a new ansatz more suitable for the description of static correlation, and where the occupied-virtual blocks of the 1RDM can be guaranteed to be zero in a future publication.

\section{Supplementary Material}
See supplementary material for the energies, cluster operator norms, amplitude errors, and partial trace defects for \ce{H2} computed by various unitary methods; the optimized diatomic geometries by various unitary methods; an explicit demonstration of the role of the matrix $U$ in DCT's construction of the 1RDM; and an explicit computation of the lowest order $d^i_a$ terms in the OUDCT ansatz.

	\begin{acknowledgments}
	We acknowledge support from the National Science Foundation, Grant No. CHE-1661604. We acknowledge helpful discussions with Professor Francesco Evangelista and Dr. Chenyang Li on iterative triples approximations in unitary theories, implementations of exact unitary coupled cluster theories, and the ability of truncated unitary coupled cluster to model static correlation.
	\end{acknowledgments}

	\appendix

\section{Analysis of the $\kappa$ and $\tau$ Decomposition of the 1RDM in DCT}
\label{sec:kappa-tau}
Many earlier DCT papers express the energy as a functional of an idempotent part of the 1RDM, $\kappa$, and the cumulant $\lambda$,\cite{Kutzelnigg:2006p171101, Simmonett_2010, Sokolov:2012p054105, Sokolov:2013p024107, Sokolov:2013p204110, Copan:2014p2389, Wang:2016p4833} with $\kappa$ said to be independent of $\lambda$.\cite{Kutzelnigg:2006p171101, Simmonett_2010, Copan:2014p2389, Sokolov:2014p074111, Wang:2016p4833} Previously reported pure $n$-representability constraints on the arguments of this functional were incomplete. We first derive the constraints and then analyze previous DCT work in terms of the complete $n$-representability constraints within this formalism.

The $\kappa, \lambda$ formalism decomposes $\gamma$ as $\kappa + \tau$. $\kappa$ is defined\cite{Kutzelnigg:2006p171101, Sokolov:2013p024107} to be the ``best idempotent approximation'' to $\gamma$.\cite{osti_4771839} That is, $\kappa$ for an $n$-electron wavefunction is the 1RDM of a Slater determinant, and its occupied orbitals are the wavefunction's $n$ natural spin-orbitals with the highest occupation numbers. $\kappa$, $\tau$, and the cumulant partial trace $d$ then have a common eigenbasis of the natural spin-orbitals. $\tau$ can be determined by 

\begin{equation}
	\label{eq:kt}
	\tau_{p^\prime} = \frac{1 \pm \sqrt{1+4\Delta_{p^\prime}}}{2} - \kappa_{p^\prime}
\end{equation}

\noindent where $p$ indexes the eigenvectors, and quantities in the natural spin-orbital basis are denoted with primed indices. $\kappa_{p^\prime}$ is 1 for an occupied natural spin-orbital and 0 otherwise, and $\Delta_{p^\prime}$ refers to a $d$ eigenvalue. Assuming that there are $n$ natural spin-orbitals with occupation number $\geq 0.5$ and all others have occupation number $\leq 0.5$, \eqref{eq:kt} simplifies to:

\begin{equation}
	\tau_{i^\prime} = \frac{-1 + \sqrt{1+4\Delta_{i^\prime}}}{2}
\end{equation}
\begin{equation}
	\tau_{a^\prime} = \frac{1 - \sqrt{1+4\Delta_{a^\prime}}}{2}
\end{equation}

Here, $\tau$ depends on both $\kappa$ and $\lambda$ by \eqref{eq:kt} and \eqref{eq:def-d}, respectively. In every case, $\kappa$ prescribes when to use the $+$ solution of \eqref{eq:d-to-gamma-eval} and when to use the $-$ solution. $\kappa$ also prescribes how to resolve degenerate eigenvectors in the $d$ matrix, if any, into occupied and virtual orbitals.

Using \eqref{eq:cumulant-energy}, the energy is then expressed as a functional of $\kappa$ and $\lambda$:\cite{Kutzelnigg:2006p171101}

\begin{multline}
	\label{eq:kappa-lambda}
	E(\kappa, \lambda) = h^q_p (\kappa^p_q + \tau^p_q(\kappa, \lambda))
	\\+ \frac{1}{2} \bar{g}^{rs}_{pq} (\kappa^p_r + \tau^p_r(\kappa, \lambda)) (\kappa^q_s + \tau^q_s(\kappa, \lambda)) +
	\frac{1}{4} \bar{g}^{rs}_{pq} \lambda^{pq}_{rs}
\end{multline}

It is now necessary to consider what constraints must be placed on $\kappa$ and $\lambda$. Some are already known.
\begin{enumerate}
	\item $\lambda$ must be derived from some wavefunction, that is, be pure $n$-representable.\cite{Kutzelnigg:2006p171101}
	\item $\kappa$ must be derived from some wavefunction, that is, be an idempotent density matrix with trace $n$.\cite{Kutzelnigg:2006p171101}
\end{enumerate}
\noindent However, the following have not been previously reported:
\begin{enumerate}[resume]
	\item \eqref{eq:kappa-lambda} is only defined for pairs of $\kappa, \lambda$ derived from \textit{the same} wavefunction. This is a stronger requirement than the above. For example, it was already known\cite{Kutzelnigg:2006p171101} that $\kappa$ and $d$ must share an eigenbasis to satisfy this condition. This has been recognized as crucial in \eqref{eq:kt}, but has not been recognized as an \textit{additional} constraint on the parameters of the energy functional.
	\item If $\kappa$ and $d$ have a common eigenbasis, then contrary to previous reports,\cite{Kutzelnigg:2006p171101, Simmonett_2010, Copan:2014p2389, Sokolov:2014p074111, Wang:2016p4833} $\kappa$ and $\lambda$ are not independent. The set of $\lambda$ allowed for a given $\kappa$ depends on $\kappa$, and likewise, the set of $\kappa$ allowed for a given $\lambda$ depends on $\lambda$.
\end{enumerate}

\noindent Insisting that \eqref{eq:kappa-lambda} is variationally minimized with respect to variations of $\kappa$ unaccompanied by $\lambda$ variations produces the DCT $\kappa$ stationarity equation of Equation 22 from Reference \citenum{Kutzelnigg:2006p171101}.

No DCT numerical studies varied $\kappa$ in this unphysical way.\cite{Simmonett_2010, Sokolov:2012p054105, Sokolov:2013p024107, Sokolov:2013p204110, Copan:2014p2389, Sokolov:2014p074111, Mullinax_2015, Wang:2016p4833, Copan:2018p4097, Peng:2019p1840} In all cases, all variations of $\kappa$ were coupled to variations of $\lambda$ so that $\kappa$ and $d$ computed from $\lambda$ had a common eigenbasis, consistent with this constraint. To explain the matter, it is convenient to change the variables of \eqref{eq:kappa-lambda}.

First, define:
\begin{equation}
	\label{eq:def-kappa-prime}
	\kappa^p_q(U) = (U^\dagger)^p_{p^\prime} (\kappa^\prime)^{p^\prime}_{q^\prime} U_q^{q^\prime}
\end{equation}

\noindent where $\kappa^\prime$ is an arbitrary but fixed idempotent density matrix with trace $n$, and $U$ is a unitary transformation. Then varying $\kappa$ so that it remains an idempotent density matrix with trace $n$ is equivalent to varying $U$ subject to it remaining unitary. $\kappa^\prime$ has always been chosen to be diagonal, so its orbitals do not mix occupied and virtual natural spin-orbitals. Choosing the primed indices to be natural spin-orbitals is consistent with this but not required.

Second, define:\cite{Sokolov:2013p204110}
\begin{equation}
	\label{eq:lambda-prime}
	\lambda^{pq}_{rs}(U, \lambda^\prime) = (U^\dagger)^p_{p^\prime} (U^\dagger)^q_{q^\prime} (\lambda^\prime)^{p^\prime q^\prime}_{r^\prime s^\prime} U_r^{r^\prime} U_s^{s^\prime} \quad .
\end{equation}

\noindent Any wavefunction with cumulant $\lambda$ will give cumulant $\lambda^\prime$ after the orbital rotation specified by the matrix $U$, and vice versa. It follows that for unitary $U$, $\lambda$ is pure $n$-representable if and only if $\lambda^\prime$ is.

Using this, we define a new energy functional:

\begin{equation}
	\label{eq:U-lambda}
	E(U, \lambda^\prime) = E(\kappa(U), \lambda(U, \lambda^\prime))
\end{equation}

\noindent These two functionals are related by a change of variables. While they have different functional dependence on their variables, their physical content is the same. However, $E(U, \lambda^\prime)$ follows previous DCT publications more closely.\cite{Simmonett_2010, Sokolov:2012p054105, Sokolov:2013p024107, Sokolov:2013p204110, Copan:2014p2389, Sokolov:2014p074111, Mullinax_2015, Wang:2016p4833, Copan:2018p4097, Peng:2019p1840}
They parameterized $\lambda^\prime$ as a finite polynomial in cumulant amplitudes and constructed $\lambda$ from \eqref{eq:lambda-prime}.\cite{Simmonett_2010, Sokolov:2012p054105, Sokolov:2013p024107, Sokolov:2013p204110, Sokolov:2014p074111, Copan:2014p2389, Mullinax_2015, Wang:2016p4833, Copan:2018p4097, Peng:2019p1840} They allowed for variations of $U$ that did not vary $\lambda^\prime$, where variations of $\kappa$ that did not vary $\lambda$ were not allowed. This is true of both DC\cite{Simmonett_2010} and ODC\cite{Sokolov:2013p204110} methods. The difference between the two is their $U$ stationarity condition. DC methods chose $U$ in $E(U, \lambda^\prime)$ to make the approximate $E(\kappa, \lambda)$ stationarity with respect to $\kappa$, while ODC methods chose $U$ in $E(U, \lambda^\prime)$ to make the approximate $E(U, \lambda^\prime)$ stationary with respect to $U$. Although DC methods enforce stationary with respect to $\kappa$ variations that violate $n$-representability, the rotations of $U$ used to satisfy that constraint preserve $n$-representability. Accordingly, the different orbital stationarity condition does not directly affect the $n$-representability of DC methods.

Let us consider $E(U, \lambda^\prime)$'s constraints on its arguments. We now have:

\begin{enumerate}
	\item $\lambda^\prime$ must be pure $n$-representable.
	\item $U$ must be a unitary matrix.
	\item \eqref{eq:U-lambda} is only defined for $\lambda^\prime$ that can be derived from a wavefunction that also yields the $\kappa^\prime$ appearing in \eqref{eq:def-kappa-prime}. This implies $\kappa^\prime$ and $\lambda^\prime$ have a common eigenbasis. In the typical case that $\kappa^\prime$ is chosen block-diagonal in the occupied and virtual blocks, this means the same must be true of $d$.
	\item The set of admissible $\lambda^\prime$ is independent of $U$. It does depend on $\kappa^\prime$, but $\kappa^\prime$ does not vary.
\end{enumerate}
\noindent All previous DCT numerical studies\cite{Simmonett_2010, Sokolov:2012p054105, Sokolov:2013p024107, Sokolov:2013p204110, Sokolov:2014p074111, Copan:2014p2389, Mullinax_2015, Wang:2016p4833, Copan:2018p4097, Peng:2019p1840} parameterized $\lambda^\prime$ with block-diagonal $d$, although this was not mentioned as a necessary constraint for \eqref{eq:U-lambda}.

As discussed in \ref{sec:oo}, the general OUDCT ansatz of Reference \citenum{Sokolov:2014p074111} does not follow this constraint of block-diagonal $d$. This does not mean that the OUDCT ansatz is inconsistent, only that its orbital rotation is to the \textit{energy minimizing orbitals} rather than the \textit{natural orbitals} from which $\kappa$ is constructed. Its energy functional is not obtained simply by parameterizing the $\lambda^\prime$ in $\eqref{eq:lambda-prime}$ used in \eqref{eq:U-lambda}, but must also modify or bypass the construction of $\kappa$ and $\tau$ in order to determine when to take $+$ or $-$ negative solutions of \eqref{eq:d-to-gamma-eval}.

	\bibliography{abbreviations.bib,JPMrefs.bib}
	
\end{document}